# Principal Stratification with Bayesian Additive Regression Trees for Count-Valued Intermediate Variables: Estimating the Effect of Fertility on Women's Employment

Lucas Godoy Garraza (lgodoygarraz@umass.edu) and Leontine Alkema

Department of Biostatistics and Epidemiology, University of Massachusetts Amherst

**Abstract**

Estimating the causal effect of fertility on women's employment is challenging because fertility and labour-market decisions are jointly determined. Instrumental-variable strategies are widely used, but credible instruments are rare and their validity often depends on covariates. Two-stage least squares — the dominant implementation — does not flexibly accommodate covariate-dependent instrument validity and is poorly suited to count-valued treatment and effect heterogeneity more broadly.

We extend an existing framework that combines principal stratification with Bayesian Additive Regression Trees (BART) to settings with count-valued intermediate variables, such as number of children. The approach defines a causal estimand that respects the count structure of the intermediate variable, while BART enables flexible, covariate-dependent modelling of principal strata and outcomes — accommodating covariate-dependent instrument validity and producing estimates of effect heterogeneity. Simulations demonstrate that our approach outperforms conventional and flexible instrumental-variable estimators under nonlinear confounding and heterogeneous treatment effects. Applied to Demographic and Health Survey data from Nigeria, Senegal, and Kenya, the approach suggests a negative average effect in Nigeria but no clear average effect elsewhere, with employment penalties concentrated among younger, less-educated women — heterogeneity that standard approaches would obscure. The approach is available in the R package PrinceBART.

# 1 Introduction

The relationship between women's labour force participation and fertility has long been of interest because it raises central questions about women's economic roles, household decision-making, and the trade-offs between work and family life (Finlay, 2021). In high-income countries, quasi-experimental studies often find that additional children reduce mothers' labour-market engagement (e.g., Angrist & Evans, 1998). In low- and middle-income countries (LMICs), the evidence is more limited and mixed (Agüero & Marks, 2008, 2011; Cáceres-Delpiano, 2012; Cruces & Galiani, 2007). Estimation is difficult because fertility and employment decisions are jointly determined and may be influenced by unobserved factors such as preferences, support networks, household bargaining power, or financial constraints. Cross-sectional associations between fertility and employment are therefore confounded and do not, on their own, reveal causal effects (Moffitt, 2005).

Quasi-experimental studies have addressed this problem using instruments that differ in the fertility margins they identify and in their suitability for low- and middle-income countries. Twin births and the sex composition of earlier children are frequently used, with Angrist & Evans (1998) providing a canonical application of this approach. These instruments are powerful but identify effects tied to specific fertility margins. Twin births combine an increase in family size with a change in the timing of births, while sex-composition instruments rely on stopping behaviour that may depend on culturally specific preferences. Agüero & Marks (2008, 2011) introduced infecundity as an alternative instrument for fertility in LMICs. Their approach is attractive because infecundity is measured in widely available and comparable Demographic and Health Survey (DHS) data and can shift parity across a broader range. Using this design, they estimated the effect of fertility on employment in 26 LMICs and found little evidence of an average effect.

Two-stage least squares (2SLS) is a standard implementation of instrumental-variable analysis and is widely used in the fertility-employment literature. In the canonical binary-treatment setting, under instrument independence, exclusion, and monotonicity, 2SLS identifies a local average treatment effect: the average causal effect among compliers, whose treatment status is shifted by the instrument (Imbens & Angrist, 1994; Angrist et al., 1996). With a multivalued ordered treatment, such as parity, 2SLS instead identifies a weighted average of per-unit causal effects over the treatment margins shifted by the instrument, often described as an average causal response (Angrist & Imbens, 1995). When instrument validity is plausible only after covariate adjustment, applied work commonly includes covariates additively in both stages. This practice is convenient, but it generally preserves a clean complier-specific interpretation only under additional functional-form restrictions (Blandhol et al., 2026; Słoczyński, 2026).

The infecundity design raises this issue directly. Because infecundity is strongly related to age and may depend on other characteristics, independence of the instrument from potential employment outcomes is unlikely to hold marginally. At best, instrument independence is plausible only after conditioning on observed covariates. In such settings, conventional 2SLS with additive covariate adjustment may fail to target a clearly interpretable complier-specific estimand. Flexible adjustment for age alone does not remove this concern when other nonlinearities remain (Blandhol et al., 2026; Słoczyński, 2026).

A further limitation is that a single average 2SLS estimate can obscure substantial heterogeneity in how an additional child affects employment across women with different ages, educational levels, or social and economic circumstances. Such heterogeneity is often explored by estimating 2SLS within subgroups, by interacting fertility with covariates, or, more recently, through machine-learning methods such as instrumental forests designed to estimate heterogeneous IV effects (Athey et al., 2019). These strategies can be informative, but they face difficulties in this setting. Subgroup and interaction-based estimates can become unstable when the instrument is weak in parts of the covariate space, a concern here because infecundity is rare at younger ages. The interpretation of subgroup and interaction-based estimates is also complicated when both the first stage and the treatment effect vary with covariates.

Several alternatives to conventional 2SLS address the covariate-dependent validity problem. Weighting approaches based on the instrument propensity score can improve adjustment for covariate-dependent instrument assignment and have been developed for binary and more general intermediate variables (Abadie, 2003; Słoczyński et al., 2025). However, these approaches are typically aimed at estimating average effects and are less directly suited to characterizing complex effect heterogeneity.

Principal stratification can simultaneously handle covariate-dependent instrument assignment and characterize effect heterogeneity. It defines causal effects within latent strata determined by how the intermediate variable would respond to the instrument. In the canonical noncompliance setting, Imbens & Rubin (1997) developed a Bayesian approach that imputes latent compliance strata and estimates stratum-specific causal effects. Recent extensions combine principal stratification with Bayesian Additive Regression Trees (BART) to flexibly model principal-stratum membership and potential outcomes in settings with binary intermediate variables (Chen et al., 2024; Godoy Garraza et al., 2026). Related work has considered continuous intermediate variables (Kim & Zigler, 2025). To our knowledge, this framework has not yet been adapted to multivalued nonnegative integer intermediate variables, such as number of children.

We extend principal stratification with Bayesian Additive Regression Trees to settings with count-valued intermediate variables. The extension is motivated by the infecundity-as-instrument design, where fertility is naturally measured as number of children and instrument validity is plausible only after adjustment for observed covariates. The proposed approach flexibly models principal-stratum membership and potential outcomes, accommodates the count-valued nature of fertility, and estimates both average and heterogeneous causal effects.

The paper is organised as follows. Section 2 introduces the empirical setting and the DHS data used in the motivating application on fertility and women's employment in low- and middle-income countries. Section 3 presents the proposed principal-stratification approach with Bayesian Additive Regression Trees for count-valued intermediate variables, including the extension to target-population estimands. Section 4 uses simulations to assess performance under nonlinear confounding and heterogeneous effects. Section 5 applies the approach to DHS data from Nigeria, Senegal, and Kenya. Section 6 concludes with a discussion of the approach, its empirical findings, and directions for future work.

# 2 Empirical setting and data

We are interested in the causal effect of fertility on women's employment in low- and middle-income countries. More specifically, we ask how a woman's probability of being employed would change if she had one additional child, and whether this effect differs across social and demographic groups. Because fertility is measured as the number of children ever born, the treatment of interest is count valued: the relevant contrast is not simply between having and not having children, but between adjacent parities.

We study this question using data from the Demographic and Health Surveys (DHS) program. DHSs are nationally representative household surveys that collect standardized information on fertility, reproductive health, employment, and demographic characteristics in low- and middle-income countries (Corsi et al., 2012). Since 1984, the DHS Program has supported more than 400 surveys in over 90 countries, making it one of the main sources of comparable population and health data for LMICs. For the empirical applications, we use the 2018 DHS in Nigeria, the 2023 DHS in Senegal, and the 2022 DHS in Kenya (National Population Commission - NPC & ICF, 2019; Agence Nationale de la Statistique et de la Démographie (ANSD) & ICF, 2024; Kenya National Bureau of Statistics et al., 2022). These countries provide three applications in sub-Saharan Africa where the DHS contains comparable measures of fertility, employment, infecundity, and demographic characteristics.

Employment is measured as a binary indicator equal to one if the respondent reports having worked for cash or in kind in the 12 months preceding the survey and zero otherwise (DHS variable v731, "Have you done any work in the last 12 months?"). Fertility is measured by parity, defined as the total number of children ever born to the respondent (DHS variable v201).

Following Agüero & Marks (2008, 2011), we use current infecundity—the reported inability to conceive—as an instrument for fertility. We classify a woman as infecund if she cites subfecundity or infecundity as her reason for not using contraception (DHS variable v3a08e). In addition, among non-sterilized women, we classify as infecund those who report an inability to have more children when asked about future childbearing desires (DHS variable v605). The identifying idea is that infecundity may reduce the number of children a woman has, while affecting employment only through fertility after conditioning on relevant covariates. This conditional qualification is important because infecundity is strongly related to age and may also vary with other characteristics.

The covariates used in the analysis are restricted to characteristics that plausibly precede infecundity, fertility, and employment. They include age in years (DHS variable v012), educational attainment (v106: no education, primary only, or some secondary and above), religion (v130: Muslim, Christian, other), urban versus rural residence (v025), and sexual experience (ever had intercourse, derived from v525). We do not adjust for variables likely to be affected by fertility or employment, such as current household wealth, partner's income, or household composition, to avoid post-treatment adjustment bias in this cross-sectional setting.

The causal estimands and notation are introduced in the next section. Briefly, the approach focuses on women whose fertility would be shifted by the infecundity instrument and estimates how the effect of an additional child varies with observed characteristics. We then summarize these effects for the affected women in the survey sample and, under additional assumptions, for broader target populations.

# 3 Methods

## 3.1 Notation and outcomes of interest

For a sample of women indexed by $i = 1, \dots, n$, $Z_i = \{0,1\}$ indicates the current infecundity status for woman $i$, $W_i(z) \in \{0,1 \dots, J\}$ the number of children a woman would have depending on her fecundity status, and $Y_i(j)$ denotes a woman's employment status if she had $j$ kids, for $j = 0, 1, \ldots, J$. We set J equal to the sample maximum plus three—placing the bound well beyond the thin upper tail of the fertility distribution—so that the support of W comfortably covers all observed values without artificially truncating.

We are interested in learning some aspect of the distribution of $Y_i(j) - Y_i\ (j-1)$, i.e., the causal effect of having an additional child on employment. We will focus on women for whom infecundity would induce having one less child than they would have had otherwise, i.e., for women with $\{i: W_i(0) = j, W_i(1) = j-1\}$, which we will refer to as affected by the instrument; this is the analogue, for the j to j-1 fertility margin, of the complier subpopulation underlying a local average treatment effect. We would like to estimate this effect as a function of covariate characteristics, referred to as the conditional average treatment effect among those affected ($CATE^a$):

$$CATE_j^a(x) = \mathbb{E}(Y_i(j) - Y_i(j-1) | X_i = x, W_i(0) = j, W_i(1) = j-1), \qquad (1)$$

and the average per child effect, $CATE^a(x) = \sum_j \kappa_j CATE_j^a(x)$, where $\kappa_j \equiv \frac{\sum_i 1[W_i(0)=j, W_i(1)=j-1]}{\sum_i 1[W_i(0)-W_i(1)=1]}$.

In addition to the effect among the segment of woman affected by the instrument in our sample, we would also like to generalize to broader populations. Specifically, we are interested in the average effect of having a $j$th child on some general population, for example, all women in Nigeria, defined as the population average treatment effect ($PATE$)

$$PATE_j = \mathbb{E}\big(Y_i(j) - Y_i(j-1)\big), \qquad (2)$$

for $j = 1, \ldots, J$. The $PATE_j$ can be obtained from combining the conditional average treatment effect in the population ($CATE$) with the covariate distribution in that population, i.e.,

$$PATE_j = \int CATE_j(x)\, dP_X(x), \tag{3}$$

where $P_X(x)$ is the distribution of covariates in the population of interest, and

$$CATE_j\ (x) = \ \mathbb{E}(Y_i(j) - Y_i(j-1)|X_i = x). \tag{4}$$

We can further summarize this quantity by the average fertility effect, $PATE = \sum_j w_j PATE_j$, where $w_j \geq 0$ may represent the relative frequency of the event $W_i = j$ in the population of interest, $\frac{1}{N}\sum_j 1[W_i = j]$ .

## 3.2 Assumptions

A number of assumptions are needed to guarantee identification of causal effects using the data at hand. Firstly, assumptions are needed to identify the causal effect among women in the survey sample who are affected by the instrument, referring to women for whom infecundity has caused her or would have caused her to have a smaller number of children. These assumptions are referred to as internal validity assumptions. Additional assumptions are required to generalize or transport the conclusion to a broader population of interest, i.e., to obtain the PATE, referred to as external validity assumptions. Textbox 1 summarizes both sets of validity assumptions, which are further discussed below by category.

**INTERNAL VALIDITY**

*Assumption 1 (Conditional unconfoundedness).*

$P\left(Z_i|X_i, W_i(0), W_i(1), \{\mathrm{Y_i}(j)\}_{\mathrm{j=0}}^{\mathrm{J}}\right) = P(Z_i|X_i)$, i.e., $Z_i$ was assigned independently of the potential outcomes after considering differences on observed covariates, $X_i$.

*Assumption 2 (Overlap).*

$0 < P(Z_i = 1|X_i) < 1$, or just $P(Z_i = 1|X_i) < 1$, if we focus on the effect among the treated, i.e., the assignment is not a deterministic function of the baseline covariates.

*Assumption 3 (Monotonicity).*

$\mathrm{W_i}(1) \leq \mathrm{W_i}(0)$, i.e., infecundity can only reduce the number of children.

*Assumption 4 (Exclusion restriction).*

$Y_i(W_i(z) = w) = Y_i(W_i(1 - z) = w)$ for $w = 1, \ldots, \mathrm{J}$, i.e., there is no direct effect of Z on the outcome.

*Assumption 5 (Conditional independence).*

$P(W_i(\mathrm{z})|W_i(1 - \mathrm{z}), X_i) = P(W_i(\mathrm{z})|\, X_i)$ for $z = 0, 1$, i.e., conditional on covariates, the potential number of children under each value of Z are independent.

**EXTERNAL VALIDITY**

*Assumption 6 (Conditional transportability).*

$\mathbb{E}\left(Y_i(1) - Y_i(0)\middle|X_i, \mathrm{W}(0), \mathrm{W}(1)\right) = \mathbb{E}(Y_i(1) - Y_i(0)|X_i)$, i.e., there are no unmeasured effect modifiers.

*Assumption 7 (Included support).*

$supp\left(P_X(X)\right) \subseteq supp\left(P_{X|\,\mathrm{W(0) \leq W(1)}}(X)\right)$, i.e., the combination of covariate values in the target population are also present among those affected by the instrument in the sample.

***Box 1:Identification assumptions***

### 3.2.1 Internal validity

Assumptions 1-4 correspond with the typical instrumental variable setting. For our application based on using current infecundity as the instrument for fertility and employment, we stress that the probability of $Z_i = 1$, i.e, of being infecund, is not marginally independent of the potential outcomes. Specifically, it is strongly associated with age. What we do assume is that conditional on age and possible other covariates, the timing of infecundity is unrelated with the desired number of kids (assumption 1) and that covariates do not entirely determine infecundity status (assumption 2).

Assumptions 3 and 4 (monotonicity and the exclusion restriction) can be problematic in some settings, but they appear reasonable here. Infecundity cannot increase a woman's total births —at most it prevents conceptions—so monotonicity holds true. And there is no obvious pathway by which being infecund would directly alter labour-force participation except through its effect on family size.

Unlike in the binary case, however, assumptions 1-4 are not sufficient to identify the joint distribution of $(W_i(0), W_i(1))$. Following Kim & Zigler (2025) for the continuous $W$ case, assumption 5 states that the association between $(W_i(0), W_i(1))$ is completely accounted by observable differences in $X_i$; we examine the sensitivity of the results to departures from this assumption in our setting.

Given assumption 1-5, we can identify the effect among women for whom infecundity has an impact—specifically, women for whom the instrument induced having one fewer child than they would have had otherwise, i.e., $\{i: W_i(0) = j, W_i(1) = j - 1\}$.

### 3.2.2 External validity

Assumptions 6-7 are introduced to estimate the average causal impact of fertility on labour-force participation in a defined target population. The assumptions require that there are no unmeasured effect modifiers and that the combination of covariate values in the target population is also present among those affected by the instrument in the sample. These assumptions are typically invoked to generalize or transport result from randomized control trials (RCT's) when participants are not randomly sampled from the population of interest. The assumptions are discussed in more detail in Wang & Tchetgen Tchetgen (2018) and Garraza et al. (2024), including conditions sufficient for 6. If assumption 6-7 hold true, then it follows that the conditional average treatment effect among the affected women equals the conditional effect in the target population ($CATE^a = CATE$), and we can obtain the PATE by averaging the CATE's over the distribution or the covariates in the population of interest, i.e.:

$$PATE_j \equiv \int CATE_j(x)\, dP_X(x) = \int CATE_j^a(x)\, dP_X(x), \tag{5}$$

where, as before, $P_X(x)$ is the distribution of covariates in the population of interest.

## 3.3 Statistical modelling and estimation

We introduce a statistical modelling approach to flexibly estimate outcomes of interest using survey data using machine learning methods. Building on Bayesian mixture models, first introduced by Imbens & Rubin (1997), our approach introduces Bayesian Additive Regression Trees (BART) into a mixture model set up. The use of BART in this setting was first introduced by Chen et al. (2024) and Godoy Garraza et al. (2026) for a binary intermediate outcome $W$, and by Kim & Zigler (2025) for a continuous intermediate outcome. We present a brief description here, with emphasis on our adaptation of the approach for a count-valued $W$.

### 3.3.1 Overview of mixture model and estimation process

For estimating causal effects, we are interested in the cross-classification of the potential intermediate outcomes, $(\mathrm{W_i}(0), \mathrm{W_i}(1))$. We define the principal stratum $G^*$ as the combination of (potential) parities based on infecundity status, i.e., $G_i^* = (W_i(0), W_i(1))$. Stratum membership is a latent variable, given that for a woman $i$, only one of the two potential parities is observed. We let $\mathcal{G}(z, w)$ denote the set of principal strata compatible with the observed combination of instrument value $Z$ and actual treatment $W$. For example, $\mathcal{G}(1, 2)$ refers to the set of strata compatible with observed $Z = 1$ and $W(1) = 2$, given by those cells with $W(0) \geq 2$.

Figure1 illustrates the principal-strata structure for a count-valued intermediate variable. In contrast to the binary case, there are multiple strata corresponding to different possible reductions in parity. Because $W$ is discrete and finite, the number of possible strata is finite; under monotonicity, strata in which infecundity would increase parity are ruled out.

| | W(Z = 0) = 0 | W(Z= 0) = 1 | W(Z= 0) = 2 | W(Z = 0) = 3 | ... |
|---|---|---|---|---|---|
| W(Z = 1) = 0 | Z did not change W | Z decreased W by 1 child | Z decreased W by 2 kids | Z decreased W by 3 kids | |
| W(Z = 1) = 1 | | Z did not change W | Z decreased W by 1 child | Z decreased W by 2 kids | |
| W(Z = 1) = 2 | | | Z did not change W | Z decreased W by 1 child | |
| W(Z = 1) = 3 | | | | Z did not change W | |
| ... | | | | | |

***Figure 1 Principal-strata structure for a count-valued intermediate variable.*** *Columns correspond to potential parity under fecundity,* $W_i(0)$*, and rows correspond to potential parity under infecundity,* $W_i(1)$*. Each cell represents a principal stratum defined by the pair* $(W_i(0), W_i(1))$*. Under monotonicity, infecundity cannot increase parity, so cells with* $W_i(1) > W_i(0)$ *are ruled out. Diagonal cells correspond to women whose parity would be unchanged by infecundity, while coloured off-diagonal cells below the diagonal correspond to women whose parity would be reduced. Cells highlighted in orange indicate the strata for which infecundity reduces parity by one child,* $W_i(1) = W_i(0) - 1$*.*

We use the latent stratum membership variable to rewrite the likelihood function into a mixture model set up. For each unit $i$ in the survey sample, we observe 4 random variables $\{Y_i, Z_i, W_i, X_i\}$. We assume the joint distribution of these variables is governed by a generic parameter $\theta$, with prior distribution $p(\theta)$, conditional on which the random variables for each unit are i.i.d. The likelihood of the observed data is proportional to (see Appendix I for details):

$$\prod_{i=1}^{n} P(X_i, Z_i, W_i, Y_i|\theta) \propto \prod_{i=1}^{n} \sum_{g \in \mathcal{G}(Z_i, W_i)} P(G_i^* = g|\, X_i, \theta_G) P(Y_i|G_i^* = g, Z_i, X_i, \theta_Y)\,, \qquad (6)$$

where $G^*$ refers to the latent principal strata, the combination of $W(z)$ and $W(1-z)$. This simplified expression suggests we need to specify two models: (i) a principal strata model, and (ii) an outcome model.

We use Bayesian additive regression trees for both (i) and (ii), discussed in the next section. Given the models and prior for the model parameters, we can approximate the posterior distribution of all model parameters and the causal estimands, even though $G^*$ is not observed, using a data augmentation (DA) approach. Details of the DA approach are given in Appendix I.

### 3.3.2 Bayesian Additive Regression Trees (BART)

Our goal is to introduce a flexible non-parametric modelling approach to estimate employment and potential parities. We use BART for this purpose. The model for employment outcomes is given by:

$$\mathbb{E}(Y_i \mid X_i = x, Z_i = z, W_i(0) = w_0, W_i(1) = w_1) = \Phi\left(\mathrm{bart}_z^{(Y)}(x, \hat{\mathrm{e}}(\mathrm{x}), w_0, w_1)\right), \qquad (7)$$

where $\Phi$ is the standard normal CDF, and $bart_z^{(Y)}(v)$ refers to a BART model for outcome $Y$ at each $z$, specified below. To mitigate regularization-induced confounding (Hahn et al., 2020), we include the instrument propensity score $e(x) = Pr(Z_i = 1 \mid X_i = x)$ as an additional covariate; $e(x)$ is estimated via a separate BART model (as recommended by Hahn et al., 2020).

We also use BART to model strata membership $G^*$. In summary, we model count-valued $W_i(z)$ for each $z = \{0, 1\}$ by introducing a latent unconstrained *continuous* variable $W_i^*(z)$. Details are as follows. The distribution of $G^*$ is equivalent to the joint distributions of $(W_i(0), W_i(1))$, parity values which are never jointly observed. By assumption 5, however, we can further factorize this joint distribution into two marginal distributions, i.e., for a stratum $G^*$ = g, referring to the combination $(W(0) = j_0, W(1) = j_1)$, it follows that

$$P(G_i^* = \mathrm{g} | X_i) = P(W_i(0) = j_0 | X_i) P(W_i(1) = j_1 | X_i). \qquad (8)$$

We model the marginal distribution of each $W$ using a (rounded) BART model suggested by Kowal & Canale (2020), building on Canale & Dunson (2013). The model is set up as follows.

We introduce a latent unconstrained continuous variable $W_i^*(z)$, and map it to counts via $W_i(z) = \max(0, \lfloor W_i^*(z) \rfloor)$. We pose that $W_i^*$

$$W_i^*(z) | X_i \sim N\left(E(W_i^*(z) | X_i), \sigma_{W_z}^2\right) \quad (9)$$

with

$$E(W_i^*(z), X_i = x) = \text{bart}_z^{\text{W}}(\text{x}, \hat{\text{e}}(\text{x})), \quad (10)$$

for $z = 0,1$. In other words, we model W by using a standard BART model for an underlying latent gaussian variable $W_i^*$. An equivalent specification, directly for $P(W_i(z))$ is as follows:

$$P(W_i(z) = j | X_i = x) = \Phi\left(\frac{a_{j+1} - \text{bart}_z^{\text{W}}(\text{x})}{\sigma_{\text{W}_\text{z}}}\right) - \Phi\left(\frac{a_j - \text{bart}_z^{\text{W}}(\text{x})}{\sigma_{\text{W}_\text{z}}}\right), \quad (11)$$

with $a_0 = -\infty$, and $a_j = j - 1$, for $j = 1, \dots, J$.

#### 3.3.2.1 BART specification

In a BART model, a regression function is modelled as an ensemble of regression trees. For our application, for each BART model, i.e. for each response type $r = \{W, Y\}$ and instrument level $z = \{0,1\}$, we use ensemble of $K$ trees,

$$bart_z^{(r)}(v) = \sum_{k=1}^{K} h\left(v; \theta_k^{(r,z)}\right), \quad (12)$$

where $v$ is the response-specific predictor vector—$(X_i, \hat{e}(X_i))$ for the $W$-model; and $(X_i, W_i(0), W_i(1), \hat{e}(X_i))$ for the $Y$-model. Each $\theta_k^{(r,z)}$ parameterizes a regression tree, and $h\left(v; \theta_k^{(r,z)}\right)$ returns that tree's prediction at $v$. Regularizing priors on $\theta_k^{(r,z)}$ favor shallow trees and shrink tree-level predictions toward a common value, so any single tree contributes only a small part of the fit, helping to prevent overfitting (Chipman et al. 2007, 2010, Hill et al., 2020).

Posterior samples for all tree ensembles are obtained via Bayesian backfitting. From each draw, we obtain $\left(W_i(0), W_i(1)\right)$, the implied stratum, and $Y_i\left(W_i(z)\right)$ , and then compute the target causal contrasts by averaging over draws (i.e., integrating over uncertainty in $\left(W_i(0), W_i(1)\right)$. Additional details on the model, prior, and fitting algorithms can be found in Appendix I.

### 3.3.3 Summarizing effect heterogeneity

To summarize effect heterogeneity, we aim to find subgroups of women with similar effects of fertility on employment. In other words, we would like to turn the covariate-specific CATEs from Prince BART into interpretable subgroups of women. To do so, we fit—separately for each country—a regression model to the estimated individual level causal effects, using the covariates as predictors. This approach, fitting simpler, surrogate models to the predicted values, follows a general strategy to summarize complex data-adaptive models such as BART (Molnar et al., 2020). We use a specific regression model, which we term a surrogate shallow tree, to identify combinations of predictors that define subgroups with relatively homogeneous $CATE^a$'s. Specifically, we fit a single, parsimonious regression tree to the posterior mean of the $CATE^a$ evaluated at each individual's covariate profile (i.e., at $x_i$). To ensure simplicity, we restrict the tree depth to 3 and prune nodes with fewer than 100 affected women (i.e., $\{i: W(0) - W(1) = 1\}$). This procedure is similar to that used by Logan et al., (2019) to explore treatment effect heterogeneity (see also Garraza et al., 2024).

Once the subgroups are identified, we examine effect heterogeneity by computing the average $CATE^a$ for each subgroup. Such an average is a "mixed" quantity, in the sense that it combines population parameters with the empirical distribution of covariates in the sample (Li et al., 2022). Let $\mathcal{J} \equiv \{i: X_i^K = \mathrm{k}\}$ denote a subgroup of the sample sharing one or a few characteristics, say $X^K \subseteq X$. The mixed CATE is given by

$$MCATE_j^a(\mathcal{J}) \equiv \frac{1}{\sum_{i:i\in\mathcal{J}} \pi_j(x_i)} \sum_{i:i\in\mathcal{J}} CATE_j^a(x_i)\, \pi_j(x_i). \tag{13}$$

where $\pi_j(x) \equiv P(W_i(0) = j, W_i(1) = j - 1 | X_i = x)$ is the probability of belonging to a stratum conditional on baseline characteristics. We can also average the $CATE^a$'s over the entire sample of women affected by infecundity to obtain the mixed average treatment effect among affected, $MATE_j^a \equiv \frac{1}{\sum_i \pi_j(x_i)} \sum_i CATE_j^a(x_i)\, \pi_j(x_i)$.

## 3.4 Generalizing results to a target population

To assess the broader relevance of our findings, we aim to estimate the population average treatment effect (PATE)—the average causal impact of fertility on labour-force participation in a defined target population—rather than the effect among women whose childbearing is shifted by the instrument. Under Assumptions 6–7, this reduces to learning the target population's covariate distribution, $P_X(x)$. We accomplish this with a "scaled" Bayesian bootstrap (Rubin, 1981) tailored to complex survey designs implemented in DHS. Full methodological details are given in Garraza et al 2024, building off work by Makela et al. (2018), Rao & Wu (2010), and Zangeneh & Little (2015). In summary, we draw bootstrap weights for each primary sampling unit (cluster) from an improper Dirichlet prior, scale them by the DHS survey weights, and then re-weight individual conditional average treatment effects to reconstruct a population-level effect. This approach simultaneously corrects for covariate shift between the instrument-affected subpopulation and the broader target population and propagates the extra uncertainty inherent in complex sampling.

## 3.5 Software

The approach is available in the R package PrinceBART(Godoy Garraza, 2026). The package can be downloaded from https://github.com/AlkemaLab/prince_BART. Further information is available at https://alkemalab.github.io/prince_BART/, with the count-valued W example included at https://alkemalab.github.io/prince_BART/articles/ordinal.html.

# 4 Simulations

In this section we use simulations to assess the finite-sample performance of Prince BART count extension. To reduce researcher discretion, we adopt the data-generating processes (DGPs) devised by Heiler (2022) and recently reused by Słoczyński et al. (2025) in in a similar context. Our setup follows theirs closely, except that we replace the binary treatment with a count variable — the setting our method is designed for — and dichotomize the outcome to better match our empirical application.

## 4.1 Simulation design

The DGPs are detailed in Box 2 and illustrated in Figure 1. Design A corresponds to an instrument that is marginally valid — compliance is homogeneous across covariate values, so the average effect among women whose fertility is affected by the instrument is identified without covariate adjustment. In the remaining designs, consistent estimation requires conditioning on X, as compliance varies with X. The DGP includes variation in the fertility response to the instrument, baseline employment, and the incremental effect of an additional child. In Designs A and B, treatment effect heterogeneity arises from the correlation between random variation in the incremental effect of an additional child on employment, $\varepsilon_i^1$, and random variation in fertility response to the instrument, $v_i$. In Designs C and D, the systematic component of the incremental effect also depends on X, introducing covariate-driven treatment-effect heterogeneity. In Design D, the instrument propensity score is nonlinear in X. Across all designs, $\delta \in \{0.01, 0.02, 0.05\}$ controls how close the propensity score $e(X_i)$ can get to 0 or 1, indexing the degree of covariate overlap between $Z_i = 0$ and $Z_i = 1$.

We generate the instrument, treatment, and potential outcomes using additive models with covariate-dependent systematic components and random variation terms.

*Instrument*

$$Z_i = 1\big(u_i < e(X_i)\big), \qquad e(X_i) = 1/1 + \exp(-\mu_z(X_i)\theta_0)$$

where $\theta_0 = \log\left(\frac{1-\delta}{\delta}\right)$ and $\delta \in \{0.01, 0.02, 0.05\}$ controls covariate overlap.

*Potential fertility and response to the instrument*

$$W_i^*(1) = \text{Poiss}(3), \;\; W_i^*(0) = W_i^*(1) + R_i.$$

The fertility differential $R_i$ is generated as

$$R_i = 1(\mu_R(X_i) > v_i),$$

where $v_i$ introduces random variation in fertility response to the instrument.

*Potential outcomes*

$$Y_i^*(j) = 1(L(j) > 0), \qquad j = 0, \dots, J.$$

The latent employment score $L(j)$ are generated recursively as

$$L(0) = \varepsilon_i^0; \qquad L(j) - L(j-1) = \varepsilon_i^\Delta - \mu_\Delta(X_i), \qquad j = 1, \dots, J.$$

Here $\mu_\Delta(X_i)$ is the systematic component of the incremental effect of an additional child, $\varepsilon_i^0$ introduces random variation in baseline employment, and $\varepsilon_i^\Delta$ introduces random variation in the incremental effect.

*Covariates and error terms*

$$X_i \sim \text{Unif}(0,1), \qquad u_i \sim \text{Unif}(0,1), \qquad \left(\varepsilon_i^0, \varepsilon_i^\Delta, v_i\right)' \sim \text{N}(0, \Sigma), \qquad \Sigma = \begin{pmatrix} 1 & 0 & 0 \\ 0 & 1 & .5 \\ 0 & .5 & 1 \end{pmatrix}.$$

The correlation between $\epsilon_i^\Delta$ (random variation in the incremental effect of an additional child) and $v_i$ (random variation in fertility response to the instrument) makes treatment-effect heterogeneity differ systematically across principal strata.

*Design-specific systematic components*

| Design | $\mu_R(x)$ | $\mu_\Delta(x)$ | $\mu_z(x)$ |
|---|---|---|---|
| A | $0$ | $.4$ | $2x - 1$ |
| B | $2x - 1$ | $.4$ | $2x - 1$ |
| C | $2x - 1$ | $.4 + (9(x+3)^2 - 111)/18.2$ | $2x - 1$ |
| D | $2x - 1$ | $.4 + (9(x+3)^2 - 111)/18.2$ | $x + x^2 - 1$ |

We consider two sample sizes, $N \in \{500, 1000\}$, and 400 Monte Carlo replications for each combination of a design, a value of δ, and a sample size.

***Box 2: Data-generating process***

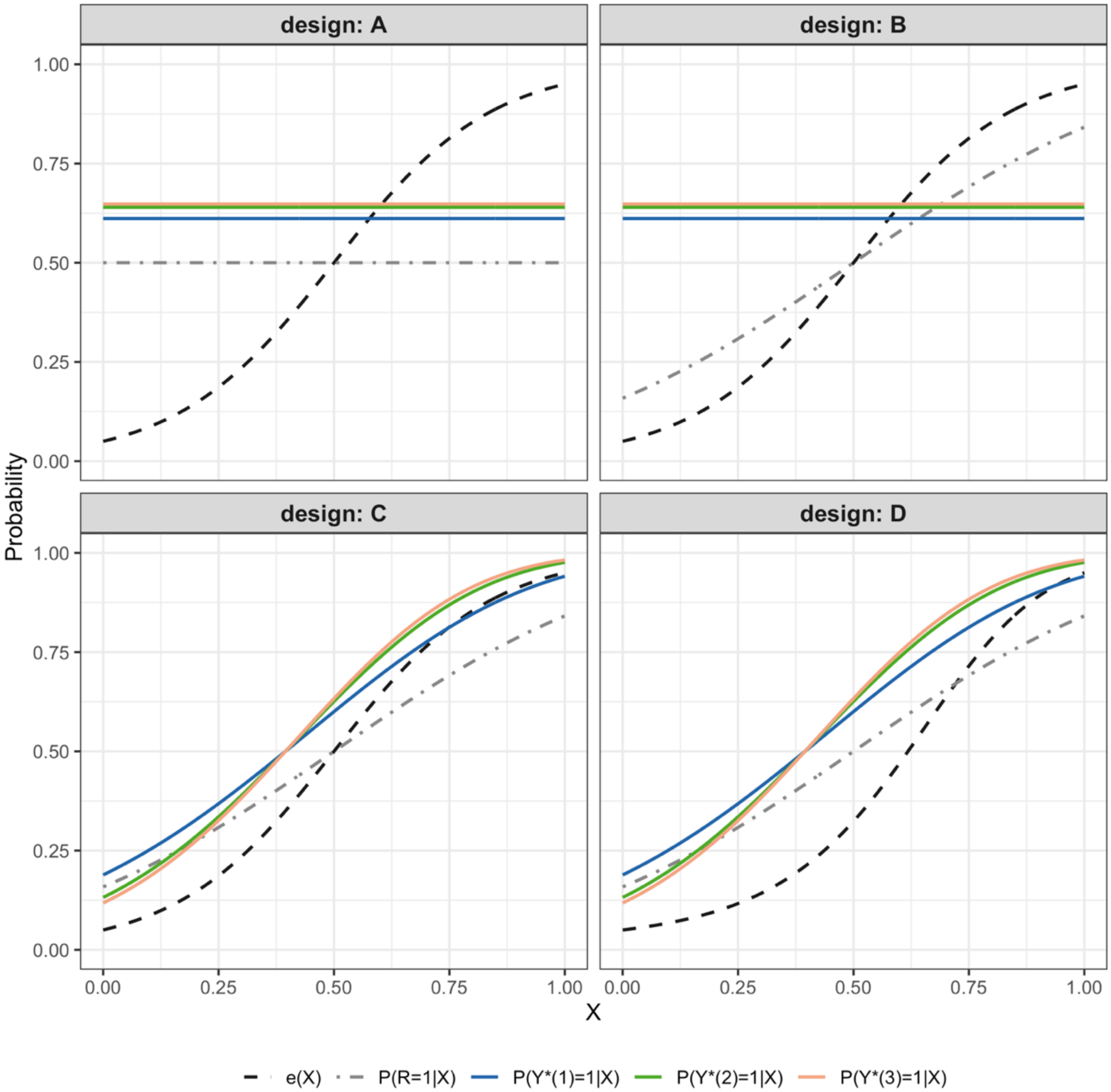

***Figure 2 Instrument, fertility-response, and employment probabilities by simulation design*** *($\delta = 0.05$). The dashed black curve is the instrument propensity score $e(X)$, the grey dot-dashed curve is the fertility-response probability $P( R = 1 \mid X )$, and the colored curves are the potential employment probabilities $P(Y(j) = 1 \mid X)$ for selected fertility levels j. Design A has homogeneous fertility response and no covariate-driven employment heterogeneity; Design B introduces covariate-dependent fertility response; Designs C and D additionally introduce covariate-driven employment heterogeneity. Design D differs from Design C by making the instrument propensity score nonlinear in X, on the probability scale shown here, this produces only a modest visual difference.*

We compare Prince BART to a set of IV estimators that vary in how they adjust for X and handle instrument propensity imbalance. An overview, including implementation details, is provided in Table 1. The main comparisons include three 2SLS specifications: an additive specification with X entered linearly, an interacted specification that allows the treatment effect to vary linearly with X, and a binned specification that adjusts for X using quantile bins. We also include DML-PLIV, which uses cross-fitted machine-learning nuisance models to partial out associations between X, the outcome, treatment, and instrument before estimating a partially linear IV effect (Chernozhukov et al., 2018). Finally, we include normalized κ-weighting, which reweights observations to target the principal-strata estimand while improving covariate balance across instrument-defined groups (Heiler, 2022, Słoczyński et al., 2025). Two additional flexible specifications—a saturated binned 2SLS estimator and generalized random forest (GRF)'s instrumental forest (Athey et al., 2019)—are reported in the appendix; both were unstable under weak overlap and not competitive in the simulations.

All simulations were run in R (R Core Team, 2026).  For Prince BART, we used default BART tree priors (Chipman et al., 2010), a probit link for the binary outcome, and 200 trees. Each simulated data set was fit using 200 warm-up iterations followed by 200 posterior iterations; code to reproduce the simulations is publicly available at https://github.com/AlkemaLab/nkids/tree/main/simulation.

| Estimator | Approach | Package |
|---|---|---|
| 2SLS additive | Parametric IV, X entered linearly | ivreg (Fox et al., 2025) |
| 2SLS interacted | Parametric IV, linear X–treatment interaction | |
| 2SLS binned | Flexible parametric IV, X adjusted via quantile bins | |
| DML-PLIV | Semiparametric IV, ML nuisance estimation | DoubleML (Bach et al., 2024) |
| κ-weighting | IPW reweighting with CBPS propensity score | kappalate (Uysal et al., 2023) |
| Prince BART | Bayesian nonparametric IV | princeBART (Godoy Garraza, 2026) |

| Estimator | Approach | Package |
|---|---|---|
| 2SLS saturated binned† | Flexible parametric IV, saturated bin dummies | ivreg (Fox et al., 2025) |
| GRF instrumental forest† | Forest-based local IV | grf (Tibshirani et al., 2026) |

***Table 1: IV estimators included in the simulation study**. Note: † Reported in Appendix II only.*

## 4.2 Simulation results

Figure 3 and Figure 4 summarize main simulation findings based on relative mean squared error (MSE) and coverage, respectively. Prince BART performs best overall, with the clearest gains at the smaller sample size and in the designs where conventional IV adjustment is expected to struggle, i.e. settings where compliance varies with X, where treatment effects are heterogeneous, and when overlap is weak. As sample size increases, the gap between Prince BART and the strongest competing estimators narrows in some designs, suggesting partial convergence among methods with sufficient information. Even so, Prince BART continues to improve on the additive 2SLS benchmark and is more stable across designs. Among the remaining estimators, 2SLS-binned and DML-PLIV are the closest competitors, sometimes improving on additive 2SLS in Designs C and D, where flexible adjustment for X reduces bias. These gains are not consistent across designs, however, and neither method matches the stability of Prince BART. Other estimators perform less reliably: 2SLS-interacted offers no clear advantage over additive 2SLS, while $\kappa$-weighting has large and unstable relative MSE throughout.

Turning to coverage (Figure 4), the picture is broadly favourable across all estimators: coverage is generally at or above the nominal 95% level, with most variation reflecting mild conservatism rather than undercoverage. The only clear exception is κ-weighting in Design D at the lowest overlap level ($\delta = 0.01$), where coverage falls to around 0.92 — below nominal, though not severely so. Prince BART is consistently the most conservative, with coverage at or near 1.00 throughout.

Taken together, the results indicate that Prince BART achieves substantially lower MSE than all competitors while maintaining valid and conservative uncertainty quantification across designs.

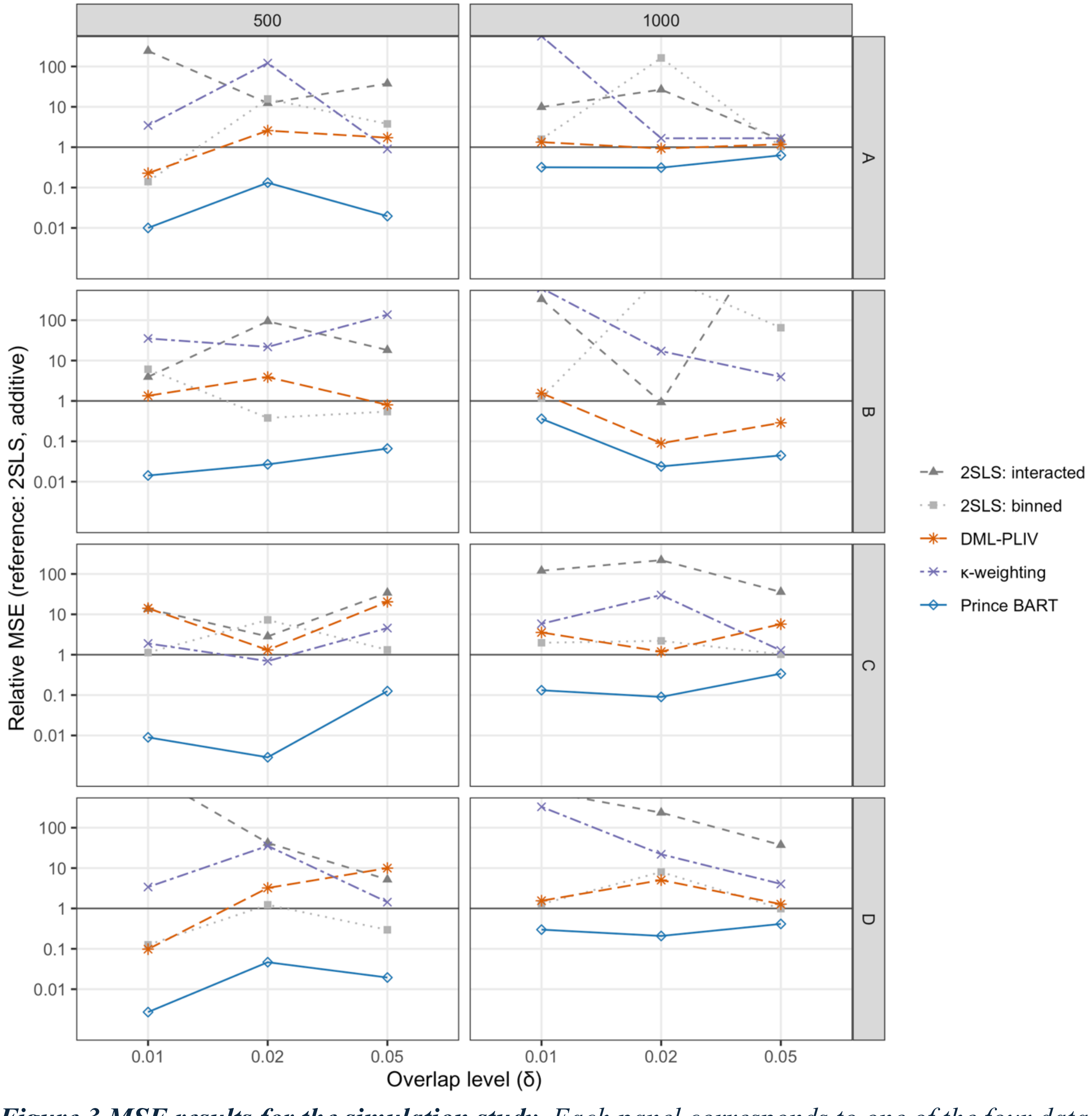

***Figure 3 MSE results for the simulation study.*** *Each panel corresponds to one of the four data-generating designs (rows) and two sample sizes (columns). The x-axis shows the overlap index δ, with higher values indicating better propensity-score overlap. The y-axis reports mean squared error (MSE) relative to the additive 2SLS estimator, so the additive 2SLS benchmark equals 1 in every panel. Values below 1 indicate smaller MSE than additive 2SLS in that scenario, while*

*values above 1 indicate larger MSE. Results are shown for interacted 2SLS, binned 2SLS, DML-PLIV, κ-weighting, and Prince BART.*

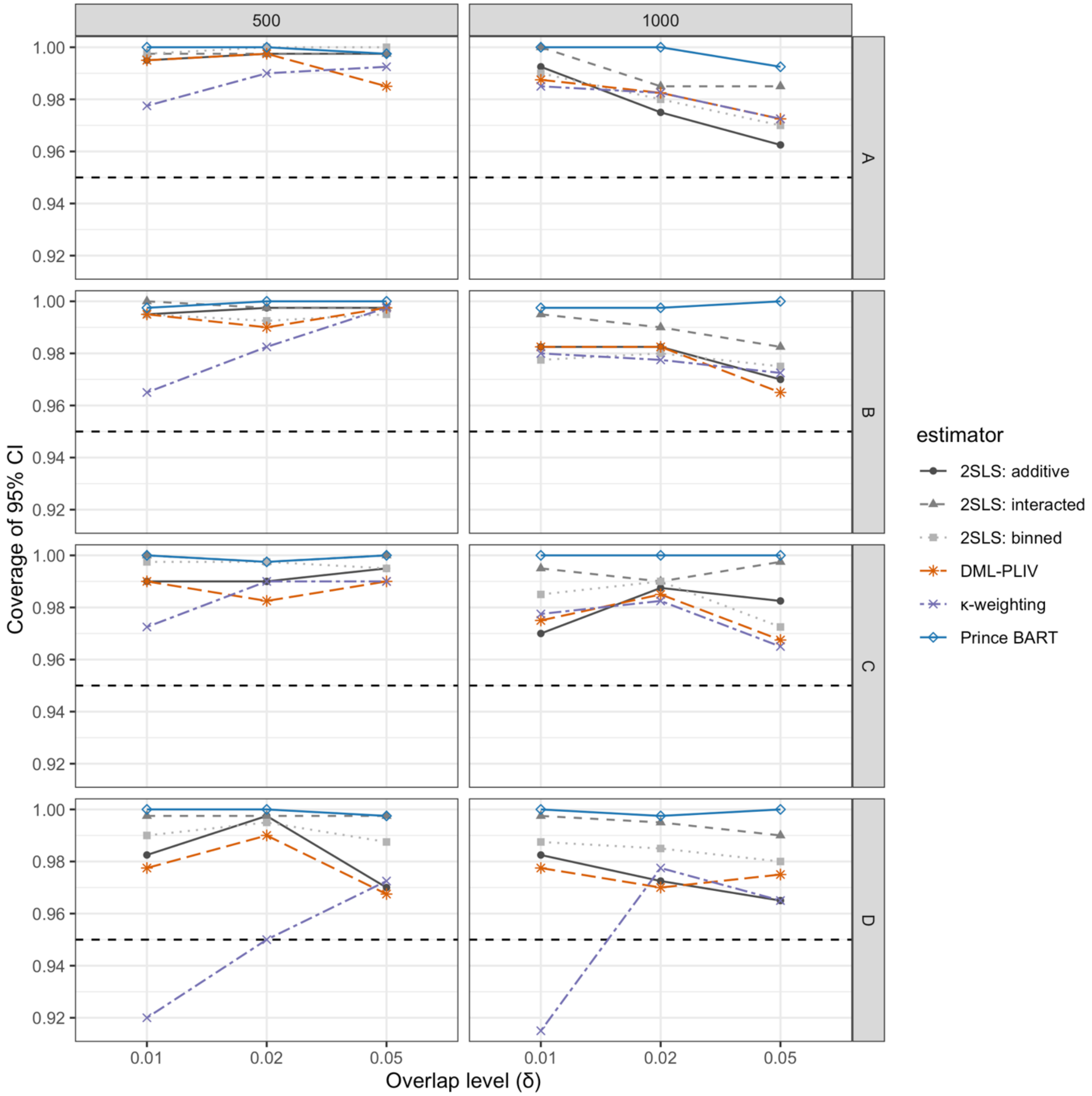

***Figure 4 Coverage results for the simulation study.*** *Each panel corresponds to one of the four data-generating designs (rows) and two sample sizes (columns). The x-axis shows the overlap index δ, with higher values indicating better propensity-score overlap. The y-axis reports empirical coverage of 95% intervals: confidence intervals for the frequentist estimators and posterior credible intervals for Prince BART. The dashed horizontal line marks the nominal 0.95*

*coverage level. Results are shown for additive 2SLS, interacted 2SLS, binned 2SLS, DML-PLIV, κ-weighting, and Prince BART.*

# 5 Application

We are interested in the causal effect of fertility on women's employment in low- and middle-income countries. As introduced in Section 2, we study this question using data from the Demographic and Health Surveys (DHS) program in Nigeria, Senegal, and Kenya.

## 5.1 Sample characteristics

Information on the samples from Kenya, Nigeria and Senegal is given in Table 1. Employment ranges from 38.35% in Senegal to 68.13% in Nigeria. Current infecundity ranges from 2.38% in Senegal to 3.05% in Nigeria. Figure 1 show the proportions of women employed ($Y$), with infecundity ($Z$) as well as the average number of kids ($W$) by age. The figure shows that all indicators typically increase with age; employment rises quickly in early adulthood before flattening out, fertility climbs steadily throughout the reproductive years, and infecundity accelerates most sharply at older ages. These distinct, non-linear age patterns underline the need for a flexible, nonparametric model of a count-valued treatment.

| | Kenya (n = 32,156) | Nigeria (n= 41,821) | Senegal (n=16,596) |
|---|---|---|---|
| Employed (%) | 54.23 | 68.13 | 38.35 |
| Infecund (%) | 2.41 | 3.05 | 2.38 |
| Parity, mean (SD) | 1.27 (2.36) | 3.37 (2.96) | 2.91 (2.57) |
| Age, mean (SD) | 29.14 (9.55) | 29.16 (9.71) | 28.09 (9.57) |
| Education: no education (%) | 11.93 | 34.43 | 46.27 |
| Education: primary only (%) | 36.72 | 15.26 | 19.60 |
| Religion: Muslim (%) | 15.09 | 50.12 | 96.79 |
| Religion: Christian (%) | 73.09 | 49.03 | 3.18 |
| Rural residence (%) | 38.52 | 40.61 | 45.05 |
| Sexual experience (%) | 83.65 | 83.77 | 73.17 |

***Table 2: Sample characteristics (unweighted).*** *Values are unweighted and describe the analytic sample used in the models. They are not intended to represent national population shares. "Employed" and "Infecund" follow our main definitions (see Methods and Appendix I). "Religion: Christian" includes Catholic and other Christian denominations; "Muslim" is separate; residual categories (other/none) are not shown. "Sexual experience" denotes ever had intercourse.*

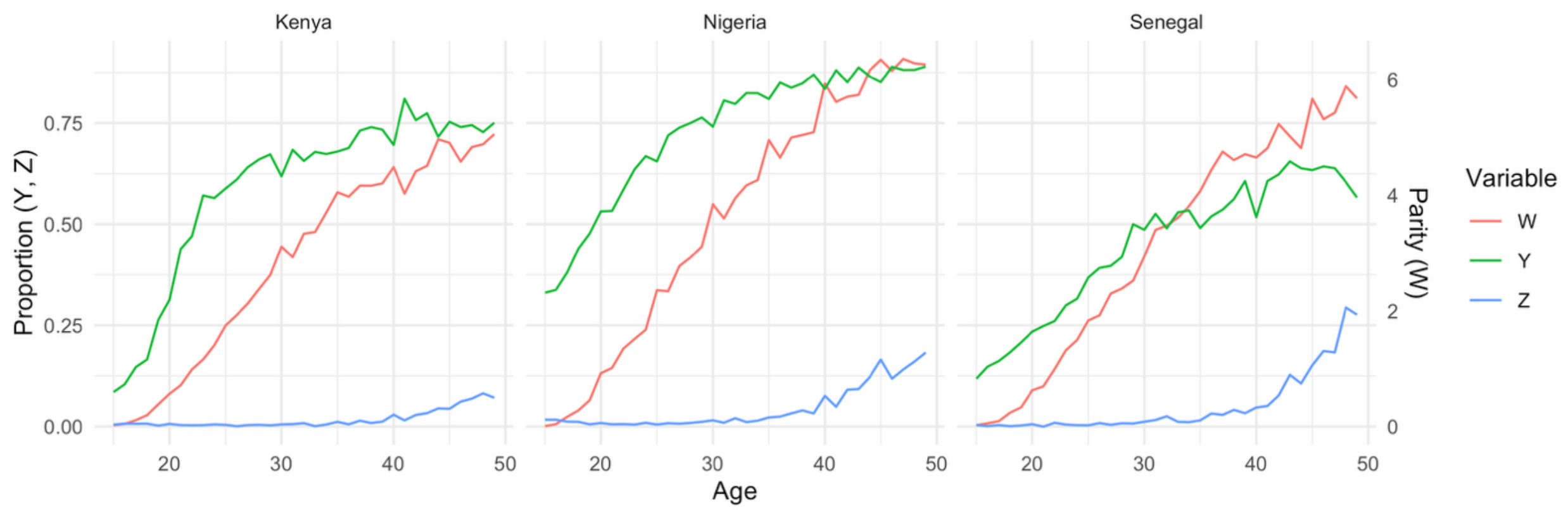

***Figure 5: Age profiles of employment, infecundity, and fertility (Kenya, Nigeria, Senegal).*** *Redline (left axis) shows the proportion employed by single-year age; green line (left axis) shows the prevalence of the infecundity instrument by age; and the blue line (right axis) plots the mean number of children ever born (fertility) by age.*

## 5.2 Average impact of an additional child

Table 2 reports the estimated impact of an additional child on the probability of employment in the past 12 months. Panel A shows the mixed-average treatment effect ($MATE^a$) among women affected by infecundity. There is evidence of a negative effect (a decrease in the probability of being employed) per additional child in Nigeria but not in Senegal or Kenya. In Nigeria, an extra child lowers employment by 6.3 percentage points (90% credible interval (CI): 1.0, 11.3). Panel B presents the population-average treatment effect, which refers to the effect among all women—not only those affected by the instrument. Findings are similar.

| Country | Panel A: MATE[a] (affected women) | | | Panel B: PATE (all women) |
|---|---|---|---|---|
| | Prince BART | | 2SLS | |
| | Main | Sensitivity Check | | |
| Nigeria | -.063 (-.113, -.010) | -.047 (-.088, -.003) | -.069 (-.102, -.037) | -.061 (-.106, -.011) |
| Senegal | .039 (-.058, .124) | .042 (-.032, .110) | .140 (-.007, .287) | .051 (-.057, .138) |
| Kenya | -.001 (-.90, .074) | -.010 (-0.76, .059) | -.048 (-.109, .013) | -.012 (-.108, .079) |

***Table 3: Estimated effect of having an additional child on the probability of being employed.*** *Columns labelled "Main" and "Sensitivity Check" report Prince BART under conditional independence and under a strong positive dependence assumption ($\rho_{(W(0),W(1))|X} = .9$), respectively. "2SLS" provides a conventional benchmark. Values in parentheses refer to 90% credible intervals for Prince BART results and 90% confidence intervals for 2SLS.*

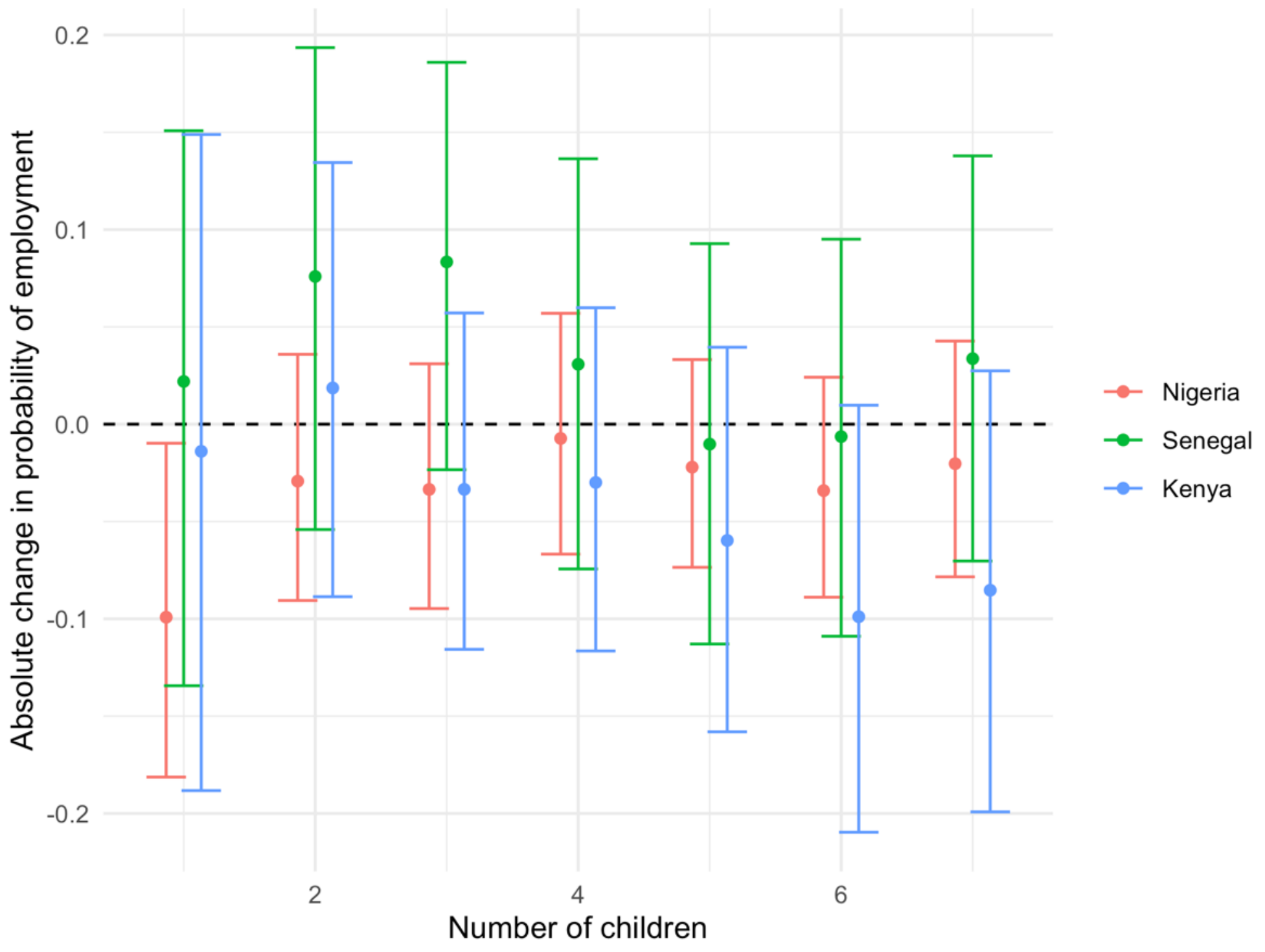

***Figure 6 Estimated effect of having an additional child on the probability of being employed ($MATE_j^a$) by number of children ($W(0)$).*** *Point estimates and 90% credible intervals are shown for each country.*

Figure 2 presents effect estimates by fertility level, i.e. the effect for each value of the intermediate value $j = W(0)$. For Nigeria, the most negative effect is at parity 1, i.e., the greatest penalty is between women with 1 versus no child. In contrast, for Kenya, effects tend to decrease with parity suggesting greater penalties at higher parity. Effects are approximately constant by parity in Senegal.

## 5.3 Effect heterogeneity

For each country, we summarize effect heterogeneity by producing estimates for different subgroups, as explained in the methods section. Figure 2 shows the posterior distribution of the difference between the largest and smallest subgroup effects in each country. The posterior probability that this difference is negative is less than 0.01 in all countries, indicating strong evidence of effect heterogeneity.

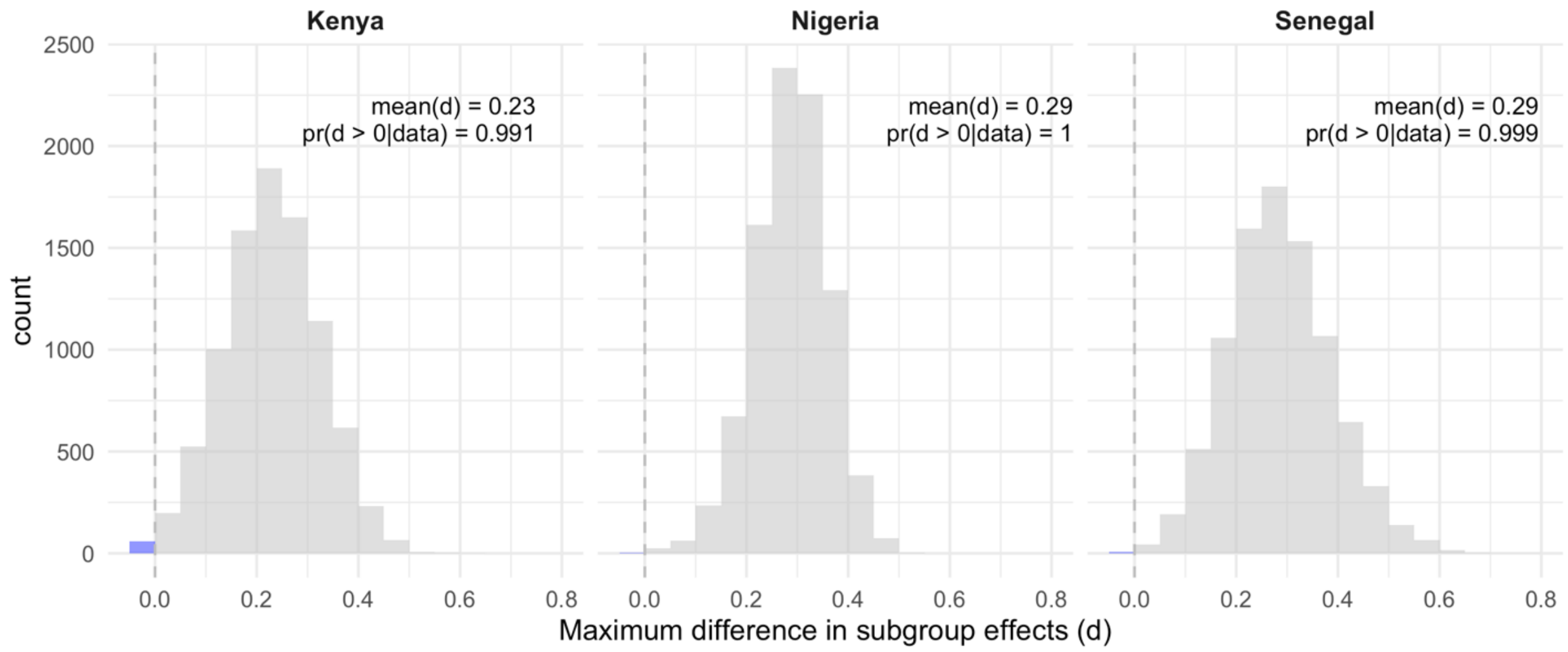

***Figure 7 Posterior distribution of the difference, d**, between the effect of an additional child on employment in the subgroups with smallest and largest effect in Kenya, Nigeria, and Senegal.*

Figures 4–6 plot the posterior mean effect of an additional child on the probability of employment (with 90% CIs) by subgroup. In Kenya, women with no education or women below

the age of 18 face the most pronounced declines (between −0.115 [90% CI: −0.223, −0.024] and −0.189 [90% CI: −0.333, −0.040]), while women over 18 years old with some education show no clear impact. In Nigeria, the sharpest employment penalty (−0.244; 90% CI: −0.327, −0.135) is concentrated among non-Muslim women younger than 20 years of age. By contrast, Muslim women older than 24 in urban areas exhibit no employment loss and may even experience a small positive effect. In Senegal, the largest negative impact (−0.139; 90% CI: −0.347, 0.030) occurs among women under the age of 20 with no education, whereas urban women aged 23–34 experience a positive effect (0.146; 90% CI: −0.001, 0.283).

As an overall summary, Figure 7 depicts the effect as a function of age, confirming that the penalty is largest for the youngest women in all three countries. We note that wider ribbons at younger ages reflect greater uncertainty—driven by a smaller share of women affected by the instrument and lower baseline employment—and affect precision, not the sign of the effect.

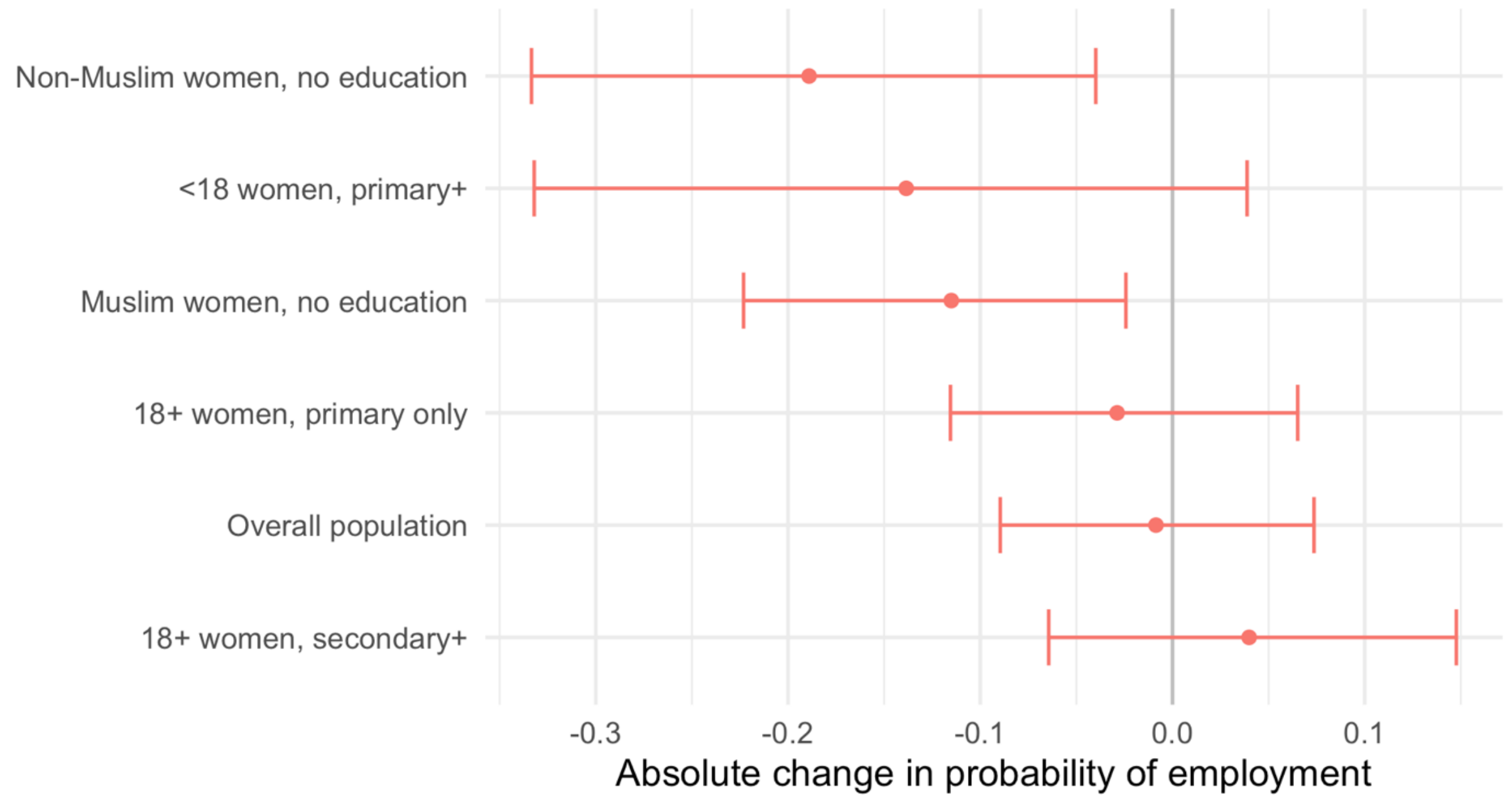

***Figure 4 Kenya***: *Employment effects by subgroup ($MCATE^{a}$). Lines represent 90% credible intervals.*

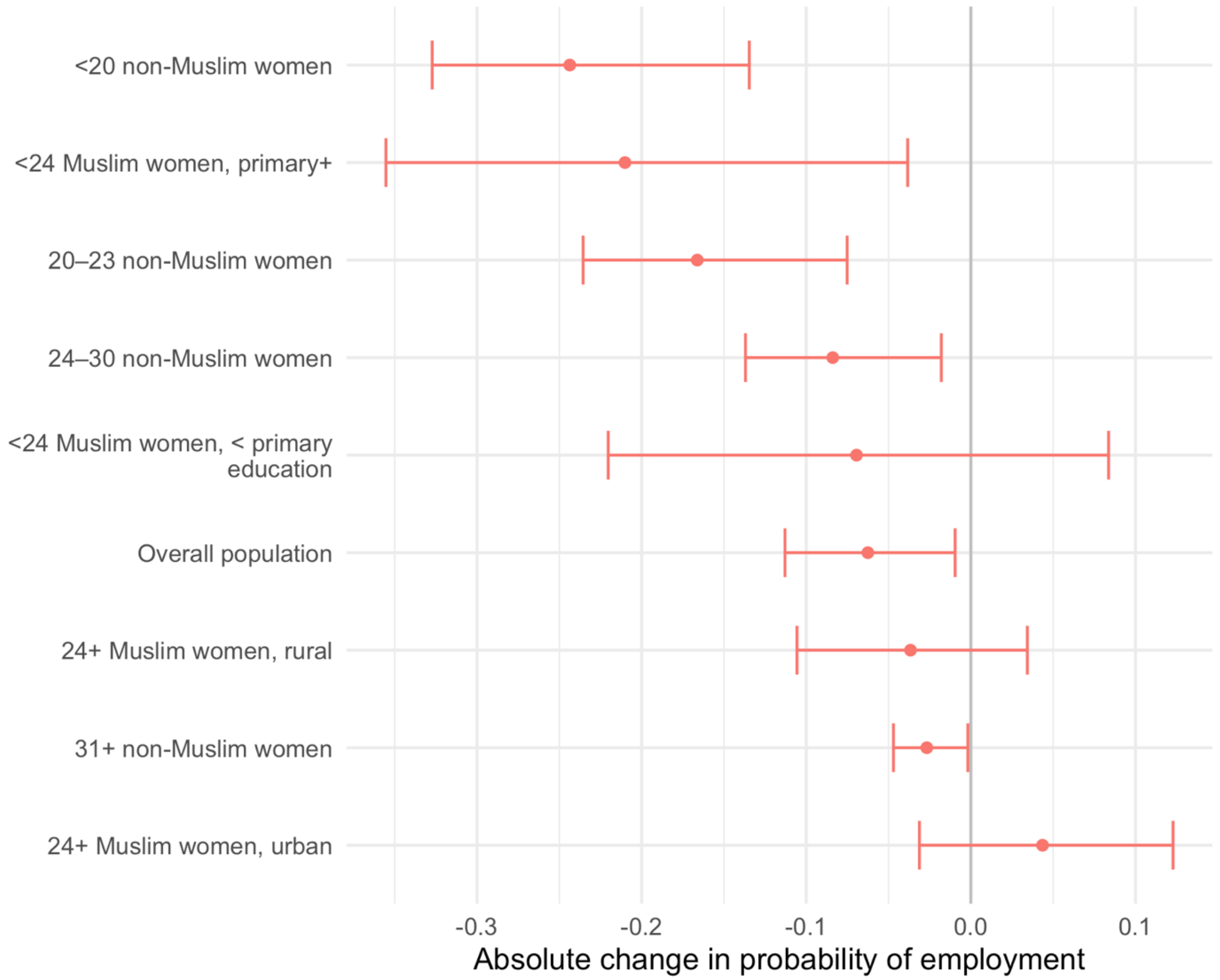

***Figure 5 Nigeria***: *Employment effects by subgroup ($MCATE^a$). Lines represent 90% credible intervals.*

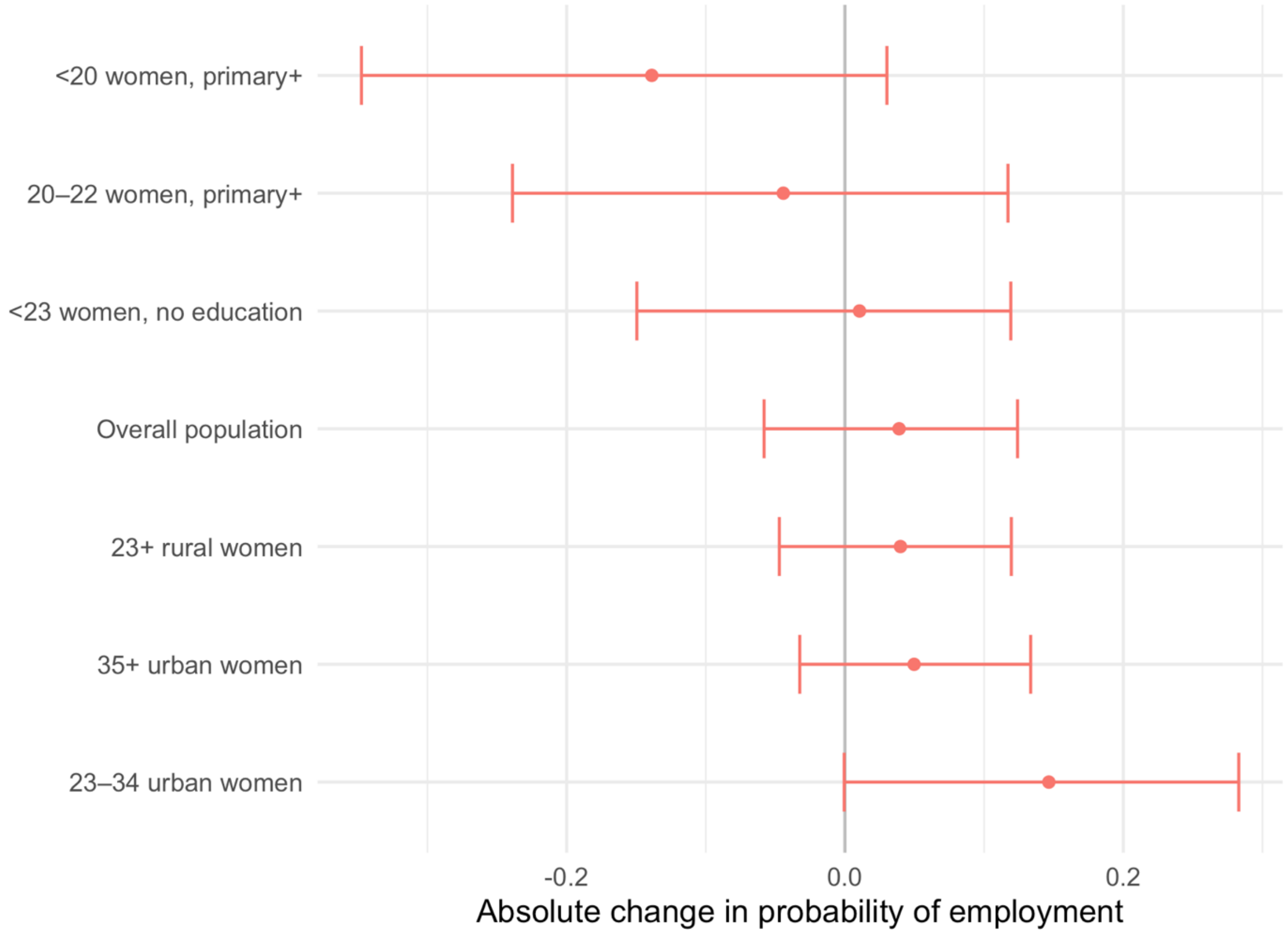

***Figure 8 Senegal***: *Employment effects by subgroup ($MCATE^{a}$). Lines represent 90% credible intervals.*

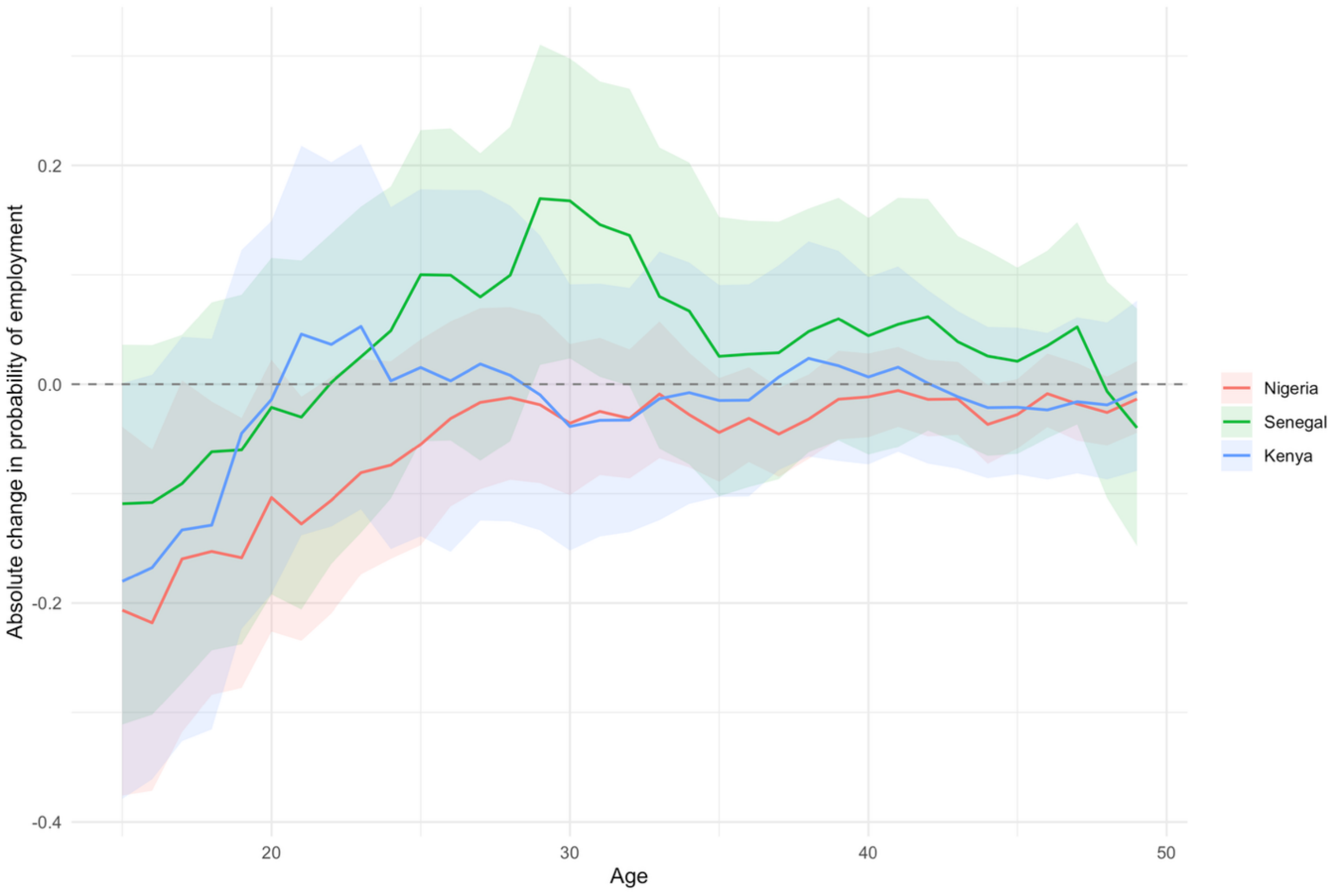

***Figure 9: Employment effects ($MCATE^a$) by age.*** *Lines show posterior means; ribbons show 90% credible intervals. Negative values indicate a decrease in the probability of employment due to fertility.*

## 5.4 Robustness checks and other comparisons

We assessed the stability of our core DHS results. Rerunning the analysis under a strong polychoric correlation assumption (0.9) between $W_i(0)$ and $W_i(1)$ we found that Prince BART's point estimates and intervals virtually unchanged, indicating robustness to violations of Assumption 5 (see Table 2). For context, Table 2 also reports a covariate-adjusted 2SLS benchmark in which age is treated as a discrete covariate, following Agüero & Marks (2008, 2011). This allows flexible adjustment for age in the first and second stages but imposes a common effect across age values. The 2SLS results are generally consistent with our Prince BART estimates: in Nigeria, both methods indicate a negative and statistically meaningful effect (e.g., −0.063 vs. −0.047), whereas in Senegal and Kenya both methods are statistically indistinguishable from zero with wide intervals.

# 6 Discussion

We adapted the Prince BART framework—originally developed for binary intermediate variables—to accommodate a count-valued treatment. This adaptation was motivated by the application to estimate the causal effect of fertility on employment, using infecundity as an instrument. The extension allows the principal-stratification model to represent potential parity under both values of the infecundity instrument and to estimate employment effects flexibly as functions of observed covariates and latent principal strata. In simulations designed to reflect key challenges of the infecundity design—covariate-dependent instrument assignment, limited overlap, nonlinear adjustment, and treatment-effect heterogeneity—Prince BART achieved substantially lower MSE than the competing IV estimators while maintaining conservative uncertainty quantification. The simulations also showed that additive 2SLS can remain a strong benchmark in simpler settings, but that more flexible IV specifications did not provide uniformly reliable improvements. In the DHS applications, the approach produced stable estimates under sensitivity checks for dependence between potential parities and allowed us to summarize both average effects and effect heterogeneity.

Our application in Kenya, Nigeria, and Senegal reveals three main substantive findings. First, having an additional child significantly reduces the probability of employment in Nigeria, but we find no clear overall employment effect in Senegal or Kenya. 2SLS results are generally consistent with Prince BART. This suggests that, for this empirical application and average estimand, the 2SLS specification does not materially change the conclusion, but the simulations caution against treating such specifications as generally innocuous.

Second, in all three countries, we find strong evidence of substantive effect heterogeneity; the posterior probability that the range of subgroup effects is zero is less than 1% in all three countries. In each setting, the employment penalty of an extra child is concentrated among younger, less-advantaged women—particularly those with lower education—while older or more educated women often see little to no negative impact. A plausible explanation is that younger, less-educated women have weaker labour-market attachment, fewer job protections, and thinner childcare or kin support, so an extra child more readily displaces work; older or more-educated women may be better able to buffer the same shock. We do not test these mechanisms here and leave their evaluation to future work.

Third, re-weighting the subgroup effects to reflect the covariate distribution of the full population produces only minor changes. This suggests that, although high- and low-effect subgroups differ somewhat in prevalence between women affected by infecundity and all women, these differences do not materially alter the estimated overall impact of fertility on employment. Our robustness checks, which impose strong residual correlation between the two potential intermediate outcomes, also leave the main conclusions intact.

In summary for the application, by combining principal stratification with Bayesian nonparametric regression, we have shown that the employment cost of an additional child is far from uniform across subpopulations and national contexts. In the present application, the clearest average penalty appears in Nigeria, but the most consistent finding across countries is heterogeneity: the costs of additional childbearing are concentrated among younger and less-educated women.

The findings are subject to important constraints. Most importantly, the analysis uses cross-sectional DHS data and depends on conditional unconfoundedness of the infecundity instrument. Infecundity is strongly related to age and may also be associated with other characteristics that affect employment. We adjust only for covariates that plausibly precede infecundity, fertility, and employment; many other DHS variables, such as household wealth, assets, or household composition, may themselves be affected by fertility or women's employment. Including such variables could introduce post-treatment bias but omitting them leaves open the possibility of residual confounding. Longitudinal data with richer pre-treatment covariates would provide a stronger basis for evaluating the identifying assumptions.

Methodologically, our contribution is a targeted extension of Prince BART to count-valued intermediate variables. By extending Bayesian nonparametric principal stratification to a count-valued treatment, we bridge the gap between existing binary and continuous implementations and provide a way to estimate incremental, parity-specific effects within a flexible Bayesian framework. The simulation study suggests that this flexibility has practical value in settings where instrument validity is conditional on covariates and where both compliance and treatment effects vary with observed characteristics. In such settings, Prince BART avoids reliance on a single linear IV specification, accommodates nonlinear covariate adjustment, and produces posterior summaries of heterogeneity while maintaining conservative interval coverage. These advantages offer demographers and applied causal-inference researchers a useful tool for studying count-valued exposures such as fertility.

# 7 Acknowledgments

The authors would like to thank Ilene Speizer, Jocelyn Finlay, Jonathan Bearak, Matt Hamilton, Onikepe Owolabi, Jennifer Seager, and John Stover, for helpful discussion and comments on this work. This paper is a product of the investigator's work within the Family Planning Impact Consortium: a multi-disciplinary partnership between the Guttmacher Institute, African Institute for Development Policy, Avenir Health, Institute for Disease Modeling of the Gates Foundation's Global Health Division, with investigators at the University of Massachusetts Amherst, the University of North Carolina, the George Washington University, and the Institut Supérieur des Sciences de la Population de l'Université Joseph Ki-Zerbo. The Consortium seeks to generate robust estimates of how family planning affects a range of social and economic domains across the life course. Members of the Consortium have developed unique model-based approaches to generating evidence that examines relationships between family planning and empowerment-related variables. AI-assisted tools were used for language editing and drafting support; the authors reviewed and take full responsibility for the submitted manuscript.

## 8 Funding

This paper was made possible by grants from the Gates Foundation and the Children's Investment Fund Foundation, who support the work of the Family Planning Impact Consortium. The findings and conclusions contained within do not necessarily reflect the positions or policies of the donors. Additional Funder information: Funder: Gates Foundation, Award Number: INV-018349, Grant Recipient: Guttmacher Institute; Funder: Children's Investment Fund Foundation, Award Number: 2012-05769, Grant Recipient: Guttmacher Institute.

# Supplemental Material for Principal Stratification with Bayesian Additive Regression Trees for Count-Valued Intermediate Variables: Estimating the Effect of Fertility on Women's Employment

Lucas Godoy Garraza (lgodoygarraz@umass.edu) and Leontine Alkema

Department of Biostatistics and Epidemiology, University of Massachusetts Amherst

# Appendix I: Prince BART for count-valued intermediate variable

The general Bayesian mixture model approach to principal stratification was introduced by Imbens & Rubin (1997). The use of BART within this framework has been extended to binary intermediate outcomes (Chen et al., 2024; Garraza et al., 2024) and to continuous intermediate outcomes (Kim & Zigler, 2025) . In this appendix, we summarize the approach emphasizing the adaptation to handle a count-valued intermediate variable, such as parity. For the implementation of the algorithm, we make extensive use of dbarts (Dorie et al., 2024) a discrete sampler that facilitates incorporating BART into more complex models.

## A1. Overview of the mixture Model

For each unit in the sample, we observe a realization of four random variables $\{X_i, Z_i, W_i, Y_i\}$. We assume their joint distribution is governed by a generic parameter vector $\theta$, with prior distribution $p(\theta)$, and that the observations are i.i.d. given $\theta$. This formulation is general, as we do not restrict the dimensionality of $\theta$. Let $G_i^* = \left(W_i(0), W_i(1)\right)$ denote the latent principal stratum—the cross-classification of the potential values of treatment $W$—and $\mathcal{G}(z, w)$ the set of principal strata compatible with the observed combination of instrument value z and actual treatment $W$. For example, $\mathcal{G}(1, 2)$ with observed $Z = 1, W(1) = 2$ has strata defined by $W(0) = 0,1,2$. Because W is discrete and finite, so is the number of different strata. Further, under the monotonicity assumption, several strata can be ruled out. Nevertheless, strata are much more numerous than in the binary case and there is a mixture of strata for almost all $(z, w)$ combinations.

The likelihood of the observed data can be written as:

$$\prod_{i=1}^{n} P(X_i, Z_i, W_i, Y_i|\theta)$$

$$= \prod_{i=1}^{n} P(X_i|\theta_X)\, P(Z_i|X_i, \theta_Z) \sum_{g\in\mathcal{G}(Z_i,W_i)} P(G_i^* = g|Z_i, X_i, \theta_G)\, P(W_i|G_i^* = g, Z_i, X_i, \theta_W)\, P(Y_i|G_i^* = g, Z_i, X_i, \theta_Y) \quad \text{(A1)}$$

$$\propto \prod_{i=1}^{n} P(X_i|\theta_X) \sum_{g\in\mathcal{G}(Z_i,W_i)} P(G_i^* = g|\, X_i, \theta_G) P(Y_i|G_i^* = g, Z_i, X_i, \theta_Y)$$

Several components are absorbed into the proportionality constant:
(i) the instrument propensity $P(Z \mid X)$ since, given Z, it has no additional information for the estimands; and

(ii) the conditional density for the parity W, since, given Z and X, W is a deterministic function of $G^*$.

Furthermore, the marginal covariate distribution $P(X)$, is not involved in the estimation of sample-based estimands.

The simplified likelihood shows that for most of our estimands we only need to specify two models: (i) a model for principal strata membership $P(G^*| X)$, and (ii) a model for outcomes $P(Y \mid G^*, X)$, along with prior distributions for their parameters. A data augmentation approach is used to approximate the posterior distribution of model parameters.

## A2. BART prior specification

We model unknown regression functions with Bayesian Additive Regression Trees (BART). In this subsection with describe the prior in more detail. To streamline exposition, we write $f(.)$ for a generic regression function with a BART (sum-of-trees) prior. Previously used $bart_z^{(Y)}(.)$ and $bart_z^{(W)}(.)$ are special cases. Throughout, $v$ denotes the predictor vector used in the preceding models. We follow default hyperparameter choices suggested by Chipman et al. (2007, 2010) and, for count-valued responses, Kowal & Canale (2020).

### A2.1. Sum-of-trees representation

A BART model represents a function as a sum of K shallow regression trees,

$$f(v) = \sum_{k=1}^{K} h(v;\theta_k), \qquad \theta_k = (T_k, M_k),$$

with K=200 in our application. The tree $T_k$ partitions the predictor space into leaves $R_{k\ell}$ with the leaf parameters $M_k = \{\mu_{k\ell}\}_\ell$. With this notation, each tree can be written as:

$$h(v;\theta_k) = \sum_{\ell} \mu_{k\ell}\, \mathbf{1}\{v \in R_{k\ell}\}.$$

### A2.2. Tree prior (structure and splits)

We use the standard branching-process prior: a node at depth $d$ splits with probability

$$p_{split}(d) = \alpha(1+d)^{-\beta}, \qquad \alpha \in (0,1), \qquad \beta \geq 0,$$

which favors shallow trees; we take the usual defaults $\alpha = 0.95$ and $\beta = 2$. At each split, the splitting variable is chosen (approximately) uniformly over predictors, and the cutpoint is chosen uniformly from available midpoints of observed values (with the usual categorical-feature handling by random subset splits when applicable).

### A2.3. Leaf parameter prior

Each terminal-node mean has a Gaussian prior that shrinks individual trees to be weak learners,

$$\mu_{k\ell} \sim N(\mu_0, \sigma_\mu^2),$$

calibrated so the induced prior on $f(\cdot)$ is concentrated around $\mu_0$ with modest spread. For modelling parity, $\mu_0$ is set at the observed sample mean. For modelling employment, $\mu_0$ is set at zero. We use the standard calibration

$$\sigma_\mu = \frac{\tau}{K},$$

with $\tau$ chosen so that $f(v)$ lies in a plausible range a priori (about 95% prior mass). For parity we match the observed range; for employment we target $\Phi(f(v)) \in [.001, .999]$. This ensures individual trees act as weak learners and the ensemble exhibits regularization toward a sensible baseline.

### A2.4. Likelihood layer (response-specific)

For parity, we place a scaled inverse-$\chi^2$ prior on the residual variance $\sigma_{\mathrm{W}}^2$, $\sigma_{\mathrm{W}}^2 \sim$ Inv-$\chi(\eta_0, s_0^2)$. We set $\eta_0 = 3$ and chose $s_0^2$ so that $P(\sigma_{\mathrm{W}} < s_W) = .90$, where $s_W$ is the sample standard deviation of $W$, yielding a weakly informative scale anchored to the data. For employment, under the probit link, the latent representation is $Y_i = \mathbf{1}\{Y_i^* \geq 0\}$, with $Y_i^* \sim N(f(v), 1)$. Thus, the residual variance is fixed at 1.

## A3. Posterior Computation via Data Augmentation

This section outlines a single MCMC sweep that ties together all model components. Let $\tilde{G}$ denote a completed version of $G^*$, where unobserved values $W(1-z)$ have been imputed. We alternate between:

1. Estimate the strata and outcome models using the current imputed values of $\tilde{G}$. In other words, estimate the probability densities functions for the latent strata and outcome with BART given observed values of $(X, Z, W, Y, \tilde{G})$.
2. Update $\tilde{G}$ (i.e., impute missing $W(1-z)$ values) given observed values $(X, Z, W, Y)$ and the current estimates of the probability density functions.

The first step is implemented simply as if $G^*$ was observed; taking imputed values as data, we estimate probability density functions for the strata and outcomes as functions of the covariates using standard routines. Given estimates of the probability density functions of $G^*$ and $Y|G^*, Z$, we apply Bayes rule to compute the probabilities of strata membership, conditional on all the observed data including the observed outcome. Finally, we use these probabilities to update $\tilde{G}$.

The detailed steps of the DA are as follows:

i. The algorithm needs to be initialized with some value for $\tilde{G}$ or, equivalently, for $W_i(1-z)$. We chose to set

$$\widetilde{W}_i(1-Z_i)^{(0)} \equiv \max(0, W_i - Z_i \cdot D_i + (1-Z_i) \cdot D_i)$$

where $D_i$ is a draw from a $\text{Poisson}(3)$ distribution.

For $l = 1, \dots, L$ iterations,

ii. Taking $\widetilde{W}_i(1-z_i)^{(l-1)}$ as if it were data, we fit a BART model using a standard BART backfitting algorithm to obtain a draw for the set of parameters $\left[\left\{\theta_k^{(\mathrm{Y,z})}\right\}_{k=1}^{K}\right]^{(l)}$ governing $Y_i(j)$ and compute $[\mathrm{P}(Y_i = 1|G_i^* = \mathrm{g}, W_i, X_i, Z_i)]^{(l)}$.

iii. Similarly, taking $\widetilde{W}_i(1-z)^{(l-1)}$ as if it were data, using standard BART backfitting algorithm we can we can obtain a draw for the set of parameters $\left[\left\{\theta_k^{(\mathrm{W,z})}\right\}_{k=1}^{K}, \sigma_{W,z}\right]^{(l)}$ governing $W_i(1-z)$ and compute $[\mathrm{P}(G_i^* = \mathrm{g}\,|W_i, X_i, Z_i)]^{(l)}$.

iv. The final step to update $\widetilde{G}_i$ is to compute its most likely value based on current draw of outcome probabilities, current implied strata probabilities, observed covariates and observed outcome using Bayes rule,

$$[\mathrm{P}(G_i^* = \mathrm{g}|Y_i = y_i, W_i, X_i, Z_i)]^{(l)} = \frac{[\mathrm{P}(Y_i = y_i|G_i^* = \mathrm{g}, W_i, X_i, Z_i)]^{(l)} \cdot [\mathrm{P}(G_i^* = \mathrm{g}\,|W_i, X_i, Z_i)]^{(l)}}{[P(Y_i = y_i|W_i, X_i, Z_i)]^{(l)}},$$

where $[\mathrm{P}(Y_i = y_i | W_i, X_i, Z_i)]^{(l)} = \sum_{\mathrm{g}:\mathrm{g} \in \mathcal{g}_i} [\mathrm{P}(Y_i = y_i | G_i^* = \mathrm{g}, W_i, X_i, Z_i)]^{(l)} \cdot [\mathrm{P}(G_i^* = \mathrm{g} \, | W_i, X_i, Z_i)]^{(l)}$ .

where $\mathcal{g}_i$ is the set of possible values of $G_i^*$. At this step, we leverage monotonicity (Assumption 3) which greatly reduces the size of $\mathcal{g}_i$.

v. Based on these posterior probabilities, we obtain a new value for $\widetilde{G}_i$ by imputing $W_i(1 - z)$, i.e.,

$$\widetilde{W}_i(1 - z)^{(l)} \sim Multinomial\left(\left\{[\Pr(G_i^* = \mathrm{g} | Y_i = y_i, W_i, X_i, Z_i = 1 - z)]^{(l)}\right\}_{\mathrm{g}:\mathrm{g} \in \mathcal{g}_i}\right)$$

We found that 18 chains of 2,000 iterations, discarding the first 500, and thinning afterwards to store 9,000 samples results in adequate convergence, as reflected in $\hat{R} \leq 1.03$ and effective sample size of several hundreds (Vehtari et al., 2020).

# Appendix II: Additional Simulation Study Results

Here we report detailed Monte Carlo results for all designs considered in the main text. Table A.1 summarizes the estimator specifications used in the simulations, including modelling choices that affect flexibility, stability, and the target estimand. For each of the four data-generating designs and for every combination of sample size n and overlap index $\delta$ (the minimum of the instrument propensity score and one minus its maximum, with larger $\delta$ indicating better overlap), Tables A.1–A.4 present the performance of the alternative estimators: 2SLS without covariates, additive 2SLS, interacted 2SLS, binned 2SLS, saturated binned 2SLS, DML-PLIV, normalized $\kappa$-weighting, GRF instrumental forest, and Prince BART. Rows correspond to (n,$\delta$)–measure combinations, and columns correspond to estimators. For each scenario, we report mean squared error relative to additive 2SLS, absolute bias, and empirical coverage of nominal 95% intervals. Coverage refers to confidence intervals for the frequentist estimators and posterior credible intervals for Prince BART. All quantities are based on 400 Monte Carlo replications and complement the graphical summaries of relative MSE and coverage shown in Figures 2 and 3 of the main text. Saturated binned 2SLS and GRF instrumental forest are included in the appendix for completeness but excluded from the main figures because of unstable performance in some scenarios.

The simulation evaluates all estimators with respect to $\tau^c$, the affected-unit average defined in Section 3.1. For ratio-based estimators such as 2SLS, the quantity actually identified is the average causal response (ACR),

$$ACR = \sum_j \omega_j \cdot \mathbb{E}\big(Y_i(j) - Y_i(j-1) | W_i(1) < j \leq W_i(0)\big),$$

where $\omega_j = \frac{N^{-1}\sum_i W_i(1) \leq j < W_i(0)}{\sum_j (N^{-1}\sum_i W_i(1) \leq j < W_i(0))}$ (Angrist & Imbens, 1995). The ACR weights dose-specific effects by the share of compliers crossing each threshold, whereas $\tau^c$ weights by the share of compliers with each potential untreated count. The two coincide when the instrument induces only a unit change in treatment, which is guaranteed by the DGP used here.

| Estimator label | Specification and implementation details |
|---|---|
| 2SLS no covariates | Unadjusted IV regression of *Y* on *W*, using *Z* as the instrument, with no covariate adjustment. Standard errors are HC3 robust. Consistent for LATE only under marginal instrument independence (Design A); in Designs B–D, where instrument validity is conditional on *X*, this specification is generally biased. |
| 2SLS additive | Additive 2SLS with outcome model *Y* ~ *X* + *W* and instrument set {*X*, *Z*}. Standard errors are HC3 robust. Targets LATE when the instrument is marginally or conditionally independent and functional forms are compatible; otherwise corresponds to a weighted average with data-dependent weights that need not correspond to compliers. |
| 2SLS interacted | Interacted 2SLS with outcome model *Y* ~ *scale(X)* × *W* and instrument set {*scale(X)*, *scale(X)* × *Z*}, where *scale(X)* denotes *X* centered and scaled. Standard errors are HC3 robust. Allows treatment effects to vary linearly with *X* but imposes parametric functional-form assumptions. |
| 2SLS binned | Additive 2SLS with *X* divided into 10 quantile bins. Includes bin indicators and a common treatment effect: *Y* ~ *q(X)*, where *q(X)* denotes a vector of bin indicators, with instrument set {*q(X)*, *Z*}. Standard errors are HC3 robust. Allows flexible adjustment for *X* in the instrument propensity and outcome levels but imposes a homogeneous treatment effect across bins. |
| 2SLS saturated binned | Saturated binned 2SLS allowing bin-specific treatment effects and first stages: *Y* ~ *q(X)* + *W:q(X)* + *X:q(X)*, instrumented by {*q(X)*, *Z:q(X)*, *X:q(X)*}. The implementation attempts 7, 6, …, 2 quantile bins and falls back to fewer bins if warnings, errors, or non-finite estimates occur. Effects are aggregated using empirical bin shares, with HC0 robust covariance. Included in the appendix only; the fitting procedure was difficult to stabilize in some scenarios. |
| DML-PLIV | Double machine learning partially linear IV estimator. Nuisance functions are estimated with random forests (*mlr3::regr.ranger*; 200 trees, minimum node size 2, maximum depth 5; Poisson split rule for the treatment model), using 5-fold cross-fitting with one repetition. Targets a constant, partially linear IV effect. |
| $\kappa$-weighting | Normalized κ-weighting implemented with *kappalate(Y ~ X \| W \| Z, std = "on")*, using covariate-balancing propensity scores (CBPS) to estimate the instrument propensity. Reweights observations to improve covariate balance across instrument-defined groups, targeting the affected-unit/LATE estimand. |
| GRF instrumental forest | Generalized random forest instrumental forest. The instrument propensity *ê(X)* is estimated with a 2,000-tree regression forest; the sample is trimmed to 0.1 < *ê(X)* < 0.9. The estimator is then fit with *grf::instrumental_forest(tune.parameters = "all")*, and effects are summarized with *grf::average_treatment_effect()*. Averaging over the trimmed covariate distribution rather than the complier distribution means this estimator does not target exactly the same estimand as Prince BART. Included in the appendix only; trimming reduced but did not eliminate extreme estimates in some scenarios. |
| Prince BART | Bayesian principal-stratification model fit with *prince_BART(Y ~ X \| Z \| W, uptake_type = "ordinal")*. BART components use default tree priors (Chipman et al., 2010) and a probit link for the binary outcome. Posterior inference uses 200 warm-up and 200 retained MCMC iterations per chain. The reported estimand is the average treatment effect among affected units (LATE), computed by averaging posterior draws of *Y(1) − Y(0)* over units with imputed *W(0) − W(1) = 1*. |

***Table A 1 Estimator specifications used in the simulation study****. Notes: All ratio-based estimators target a weighted average of dose-specific effects when the instrument induces more than a one-unit change in treatment. This distinction does not arise in the present simulations, where the DGP is constructed so that compliers shift treatment by exactly one unit. † Included in the appendix only. Both estimators were unstable under weak overlap. ‡ Reference estimator for relative MSE in all figures and appendix tables.*

| n | delta | measure | 2SLS no covariates | 2SLS additive | 2SLS interacted | 2SLS binned | 2SLS saturated binned | DML-PLIV | κ-weighting | GRF instrumental forest | Prince BART |
|---|---|---|---|---|---|---|---|---|---|---|---|
| 500 | 0.01 | MSE | 0.017 | 1.000 | 2.4E+02 | 0.140 | 56.330 | 0.227 | 3.465 | 5.3E+03 | 0.010 |
| 500 | 0.01 | \|Bias\| | 0.019 | 0.052 | 0.681 | 0.049 | 0.231 | 0.030 | 0.151 | 2.459 | 0.052 |
| 500 | 0.01 | Coverage | 0.988 | 0.995 | 0.998 | 0.998 | 1.000 | 0.995 | 0.978 | 0.980 | 1.000 |
| 500 | 0.02 | MSE | 0.303 | 1.000 | 12.345 | 15.655 | 1.6E+04 | 2.571 | 1.2E+02 | 2.5E+08 | 0.132 |
| 500 | 0.02 | \|Bias\| | 0.008 | 0.005 | 0.021 | 0.024 | 1.175 | 0.025 | 0.018 | 1.7E+02 | 0.041 |
| 500 | 0.02 | Coverage | 0.988 | 0.998 | 0.998 | 1.000 | 1.000 | 0.998 | 0.990 | 0.990 | 1.000 |
| 500 | 0.05 | MSE | 0.042 | 1.000 | 37.645 | 3.776 | 4.7E+03 | 1.719 | 0.903 | 2.0E+05 | 0.020 |
| 500 | 0.05 | \|Bias\| | 0.015 | 0.047 | 0.251 | 0.045 | 1.691 | 0.035 | 0.050 | 9.883 | 0.049 |
| 500 | 0.05 | Coverage | 0.983 | 0.998 | 0.998 | 1.000 | 1.000 | 0.985 | 0.993 | 0.978 | 0.998 |
| 1000 | 0.01 | MSE | 0.356 | 1.000 | 9.867 | 1.583 | 1.8E+02 | 1.335 | 5.6E+02 | 6.8E+05 | 0.319 |
| 1000 | 0.01 | \|Bias\| | 0.018 | 0.015 | 0.036 | 0.010 | 0.065 | 0.013 | 0.060 | 1.911 | 0.045 |
| 1000 | 0.01 | Coverage | 0.970 | 0.993 | 1.000 | 0.990 | 1.000 | 0.988 | 0.985 | 0.985 | 1.000 |
| 1000 | 0.02 | MSE | 0.283 | 1.000 | 26.756 | 1.6E+02 | 2.0E+03 | 0.928 | 1.662 | 2.7E+07 | 0.311 |
| 1000 | 0.02 | \|Bias\| | 0.010 | 0.003 | 0.023 | 0.077 | 0.369 | 0.004 | 0.001 | 34.075 | 0.052 |
| 1000 | 0.02 | Coverage | 0.970 | 0.975 | 0.985 | 0.980 | 0.998 | 0.983 | 0.983 | 0.985 | 1.000 |
| 1000 | 0.05 | MSE | 0.482 | 1.000 | 1.525 | 1.037 | 5.6E+04 | 1.179 | 1.662 | 1.1E+04 | 0.629 |
| 1000 | 0.05 | \|Bias\| | 0.016 | 0.023 | 0.025 | 0.023 | 0.789 | 0.022 | 0.026 | 0.428 | 0.057 |
| 1000 | 0.05 | Coverage | 0.978 | 0.963 | 0.985 | 0.970 | 1.000 | 0.973 | 0.973 | 0.983 | 0.993 |

***Table A 2 Monte Carlo performance of alternative estimators in Design A.*** *Rows correspond to combinations of sample size n, overlap index δ , and performance measure: relative MSE, absolute bias, and coverage. Columns correspond to 2SLS without covariates, additive 2SLS, interacted 2SLS, binned 2SLS, saturated binned 2SLS, DML-PLIV, normalized κ-weighting, GRF instrumental forest, and Prince BART. Relative MSE is normalized so that the additive 2SLS estimator equals 1 in each (n,δ) scenario. Coverage refers to nominal 95% confidence intervals for the frequentist estimators and 95% posterior credible intervals for Prince BART. Results are based on 400 Monte Carlo replications. Saturated binned 2SLS and GRF instrumental forest are included for completeness but excluded from the main figures because of unstable performance in some scenarios.*

| n | delta | measure | 2SLS no covariates | 2SLS additive | 2SLS interacted | 2SLS binned | 2SLS saturated binned | DML-PLIV | κ-weighting | GRF instrumental forest | Prince BART |
|---|---|---|---|---|---|---|---|---|---|---|---|
| 500 | 0.01 | MSE | 98.913 | 1.000 | 3.943 | 6.106 | 3.4E+02 | 1.345 | 35.248 | 6.0E+05 | 0.014 |
| 500 | 0.01 | \|Bias\| | 0.236 | 0.040 | 0.103 | 0.053 | 0.506 | 0.006 | 0.158 | 24.130 | 0.037 |
| 500 | 0.01 | Coverage | 0.993 | 0.995 | 1.000 | 0.995 | 0.998 | 0.995 | 0.965 | 0.990 | 0.998 |
| 500 | 0.02 | MSE | 0.351 | 1.000 | 93.971 | 0.379 | 6.4E+02 | 3.893 | 21.809 | 5.7E+05 | 0.027 |
| 500 | 0.02 | \|Bias\| | 0.018 | 0.026 | 0.322 | 0.028 | 0.913 | 0.039 | 0.165 | 15.967 | 0.046 |
| 500 | 0.02 | Coverage | 0.995 | 0.998 | 0.998 | 0.993 | 1.000 | 0.990 | 0.983 | 0.988 | 1.000 |
| 500 | 0.05 | MSE | 4.367 | 1.000 | 18.189 | 0.545 | 1.1E+02 | 0.801 | 1.4E+02 | 1.2E+06 | 0.066 |
| 500 | 0.05 | \|Bias\| | 0.043 | 0.000 | 0.029 | 0.013 | 0.161 | 0.016 | 0.155 | 13.376 | 0.044 |
| 500 | 0.05 | Coverage | 0.993 | 0.998 | 0.998 | 0.995 | 1.000 | 0.998 | 0.998 | 0.983 | 1.000 |
| 1000 | 0.01 | MSE | 0.960 | 1.000 | 3.3E+02 | 1.197 | 1.4E+02 | 1.557 | 6.2E+02 | 5.7E+05 | 0.361 |
| 1000 | 0.01 | \|Bias\| | 0.013 | 0.001 | 0.075 | 0.000 | 0.052 | 0.002 | 0.201 | 4.232 | 0.047 |
| 1000 | 0.01 | Coverage | 0.980 | 0.983 | 0.995 | 0.978 | 0.995 | 0.983 | 0.980 | 0.993 | 0.998 |
| 1000 | 0.02 | MSE | 0.073 | 1.000 | 0.928 | 1.3E+03 | 52.622 | 0.090 | 17.174 | 7.5E+04 | 0.024 |
| 1000 | 0.02 | \|Bias\| | 0.023 | 0.039 | 0.015 | 0.791 | 0.182 | 0.015 | 0.123 | 4.666 | 0.049 |
| 1000 | 0.02 | Coverage | 0.980 | 0.983 | 0.990 | 0.980 | 1.000 | 0.983 | 0.978 | 0.993 | 0.998 |
| 1000 | 0.05 | MSE | 0.236 | 1.000 | 4.7E+04 | 65.100 | 2.4E+04 | 0.288 | 3.953 | 1.3E+05 | 0.045 |
| 1000 | 0.05 | \|Bias\| | 0.013 | 0.005 | 3.580 | 0.116 | 2.468 | 0.001 | 0.025 | 8.824 | 0.047 |
| 1000 | 0.05 | Coverage | 0.975 | 0.970 | 0.983 | 0.975 | 0.998 | 0.965 | 0.973 | 0.978 | 1.000 |

***Table A 3 Monte Carlo performance of alternative estimators in Design B.*** *Rows correspond to combinations of sample size n, overlap index δ , and performance measure: relative MSE, absolute bias, and coverage. Columns correspond to 2SLS without covariates, additive 2SLS, interacted 2SLS, binned 2SLS, saturated binned 2SLS, DML-PLIV, normalized κ-weighting, GRF instrumental forest, and Prince BART. Relative MSE is normalized so that the additive 2SLS estimator equals 1 in each (n,δ) scenario.*

*Coverage refers to nominal 95% confidence intervals for the frequentist estimators and 95% posterior credible intervals for Prince BART. Results are based on 400 Monte Carlo replications. Saturated binned 2SLS and GRF instrumental forest are included for completeness but excluded from the main figures because of unstable performance in some scenarios.*

| n | delta | measure | 2SLS no covariates | 2SLS additive | 2SLS interacted | 2SLS binned | 2SLS saturated binned | DML-PLIV | κ-weighting | GRF instrumental forest | Prince BART |
|---|---|---|---|---|---|---|---|---|---|---|---|
| 500 | 0.01 | MSE | 1.2E+03 | 1.000 | 13.615 | 1.128 | 3.9E+02 | 14.148 | 1.897 | 1.3E+05 | 0.009 |
| 500 | 0.01 | \|Bias\| | 2.893 | 0.067 | 0.067 | 0.027 | 0.903 | 0.143 | 0.089 | 8.901 | 0.010 |
| 500 | 0.01 | Coverage | 0.402 | 0.990 | 1.000 | 0.998 | 1.000 | 0.990 | 0.973 | 0.988 | 1.000 |
| 500 | 0.02 | MSE | 1.3E+02 | 1.000 | 2.833 | 7.278 | 2.1E+02 | 1.307 | 0.694 | 1.4E+04 | 0.003 |
| 500 | 0.02 | \|Bias\| | 2.114 | 0.026 | 0.034 | 0.099 | 0.779 | 0.020 | 0.037 | 8.020 | 0.019 |
| 500 | 0.02 | Coverage | 0.422 | 0.990 | 0.998 | 0.998 | 1.000 | 0.983 | 0.990 | 0.983 | 0.998 |
| 500 | 0.05 | MSE | 4.0E+04 | 1.000 | 34.249 | 1.318 | 77.765 | 20.383 | 4.571 | 5.4E+06 | 0.125 |
| 500 | 0.05 | \|Bias\| | 0.166 | 0.005 | 0.017 | 0.004 | 0.000 | 0.017 | 0.004 | 6.121 | 0.027 |
| 500 | 0.05 | Coverage | 0.402 | 0.995 | 1.000 | 0.995 | 1.000 | 0.990 | 0.990 | 0.983 | 1.000 |
| 1000 | 0.01 | MSE | 1.5E+02 | 1.000 | 1.2E+02 | 1.975 | 97.001 | 3.562 | 5.829 | 1.3E+05 | 0.132 |
| 1000 | 0.01 | \|Bias\| | 1.343 | 0.025 | 0.000 | 0.008 | 0.038 | 0.005 | 0.014 | 1.440 | 0.023 |
| 1000 | 0.01 | Coverage | 0.130 | 0.970 | 0.995 | 0.985 | 1.000 | 0.975 | 0.978 | 0.988 | 1.000 |
| 1000 | 0.02 | MSE | 1.0E+02 | 1.000 | 2.2E+02 | 2.203 | 1.2E+02 | 1.190 | 30.072 | 2.6E+05 | 0.090 |
| 1000 | 0.02 | \|Bias\| | 1.202 | 0.004 | 0.083 | 0.011 | 0.047 | 0.024 | 0.036 | 2.596 | 0.026 |
| 1000 | 0.02 | Coverage | 0.105 | 0.988 | 0.990 | 0.990 | 1.000 | 0.985 | 0.983 | 0.980 | 1.000 |
| 1000 | 0.05 | MSE | 1.7E+02 | 1.000 | 35.807 | 1.024 | 2.0E+03 | 5.721 | 1.272 | 4.0E+05 | 0.339 |
| 1000 | 0.05 | \|Bias\| | 1.014 | 0.002 | 0.013 | 0.002 | 0.198 | 0.001 | 0.015 | 0.275 | 0.033 |
| 1000 | 0.05 | Coverage | 0.108 | 0.983 | 0.998 | 0.973 | 1.000 | 0.968 | 0.965 | 0.990 | 1.000 |

***Table A 4 Monte Carlo performance of alternative estimators in Design C.*** *Rows correspond to combinations of sample size n, overlap index δ , and performance measure: relative MSE, absolute bias, and coverage. Columns correspond to 2SLS without covariates, additive 2SLS, interacted 2SLS, binned 2SLS, saturated binned 2SLS, DML-PLIV, normalized κ-weighting, GRF instrumental forest, and Prince BART. Relative MSE is normalized so that the additive 2SLS estimator equals 1 in each (n,δ) scenario. Coverage refers to nominal 95% confidence intervals for the frequentist estimators and 95% posterior credible intervals for Prince*

*BART. Results are based on 400 Monte Carlo replications. Saturated binned 2SLS and GRF instrumental forest are included for completeness but excluded from the main figures because of unstable performance in some scenarios.*

| n | delta | measure | 2SLS no covariates | 2SLS additive | 2SLS interacted | 2SLS binned | 2SLS saturated binned | DML-PLIV | κ-weighting | GRF instrumental forest | Prince BART |
|---|---|---|---|---|---|---|---|---|---|---|---|
| 500 | 0.01 | MSE | 6.291 | 1.000 | 2.2E+03 | 0.126 | 20.019 | 0.098 | 3.395 | 2.9E+03 | 0.003 |
| 500 | 0.01 | \|Bias\| | 1.193 | 0.073 | 1.261 | 0.020 | 0.141 | 0.009 | 0.012 | 1.094 | 0.016 |
| 500 | 0.01 | Coverage | 0.372 | 0.983 | 0.998 | 0.990 | 0.993 | 0.978 | 0.920 | 0.995 | 1.000 |
| 500 | 0.02 | MSE | 3.6E+02 | 1.000 | 42.088 | 1.230 | 8.5E+02 | 3.218 | 35.451 | 9.3E+05 | 0.047 |
| 500 | 0.02 | \|Bias\| | 1.491 | 0.027 | 0.025 | 0.002 | 0.196 | 0.026 | 0.078 | 13.810 | 0.015 |
| 500 | 0.02 | Coverage | 0.415 | 0.998 | 0.998 | 0.995 | 0.998 | 0.990 | 0.950 | 0.998 | 1.000 |
| 500 | 0.05 | MSE | 1.1E+03 | 1.000 | 5.200 | 0.296 | 1.3E+03 | 9.979 | 1.433 | 2.2E+04 | 0.019 |
| 500 | 0.05 | \|Bias\| | 0.946 | 0.030 | 0.018 | 0.001 | 0.742 | 0.084 | 0.004 | 2.543 | 0.027 |
| 500 | 0.05 | Coverage | 0.390 | 0.970 | 0.998 | 0.988 | 1.000 | 0.968 | 0.973 | 0.980 | 0.998 |
| 1000 | 0.01 | MSE | 2.5E+02 | 1.000 | 8.1E+02 | 1.280 | 6.5E+02 | 1.551 | 3.3E+02 | 3.2E+05 | 0.299 |
| 1000 | 0.01 | \|Bias\| | 1.171 | 0.027 | 0.077 | 0.003 | 0.149 | 0.019 | 0.105 | 3.234 | 0.027 |
| 1000 | 0.01 | Coverage | 0.095 | 0.983 | 0.998 | 0.988 | 0.995 | 0.978 | 0.915 | 0.975 | 1.000 |
| 1000 | 0.02 | MSE | 2.8E+02 | 1.000 | 2.4E+02 | 8.009 | 5.7E+02 | 5.045 | 21.949 | 1.5E+05 | 0.208 |
| 1000 | 0.02 | \|Bias\| | 1.196 | 0.027 | 0.062 | 0.006 | 0.078 | 0.031 | 0.044 | 0.334 | 0.030 |
| 1000 | 0.02 | Coverage | 0.112 | 0.973 | 0.995 | 0.985 | 0.998 | 0.970 | 0.978 | 0.988 | 0.998 |
| 1000 | 0.05 | MSE | 2.7E+02 | 1.000 | 37.294 | 0.989 | 3.8E+02 | 1.276 | 4.044 | 9.4E+04 | 0.413 |
| 1000 | 0.05 | \|Bias\| | 0.914 | 0.027 | 0.014 | 0.010 | 0.042 | 0.020 | 0.031 | 0.907 | 0.036 |
| 1000 | 0.05 | Coverage | 0.085 | 0.965 | 0.990 | 0.980 | 1.000 | 0.975 | 0.965 | 0.988 | 1.000 |

***Table A 5 Monte Carlo performance of alternative estimators in Design D.*** *Rows correspond to combinations of sample size n, overlap index δ , and performance measure: relative MSE, absolute bias, and coverage. Columns correspond to 2SLS without covariates, additive 2SLS, interacted 2SLS, binned 2SLS, saturated binned 2SLS, DML-PLIV, normalized κ-weighting, GRF instrumental forest, and Prince BART. Relative MSE is normalized so that the additive 2SLS estimator equals 1 in each (n,δ) scenario. Coverage refers to nominal 95% confidence intervals for the frequentist estimators and 95% posterior credible intervals for Prince*

*BART. Results are based on 400 Monte Carlo replications. Saturated binned 2SLS and GRF instrumental forest are included for completeness but excluded from the main figures because of unstable performance in some scenarios.*

# Appendix III: Additional plots

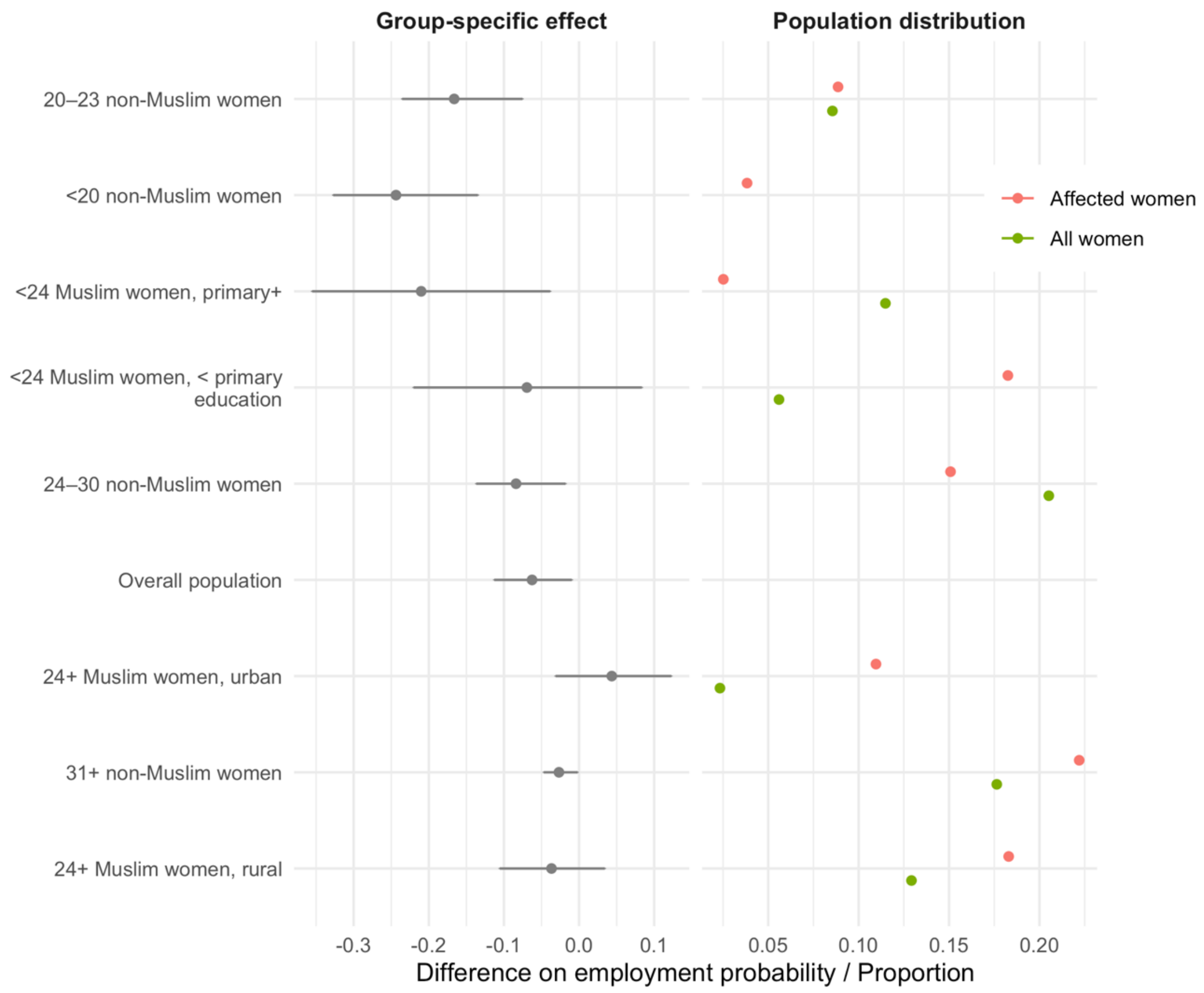

***Figure A 1 Nigeria**: Employment effects by subgroup and distribution of the subgroups among women affected by infecundity and all women. Lines represent 90% credible intervals.*

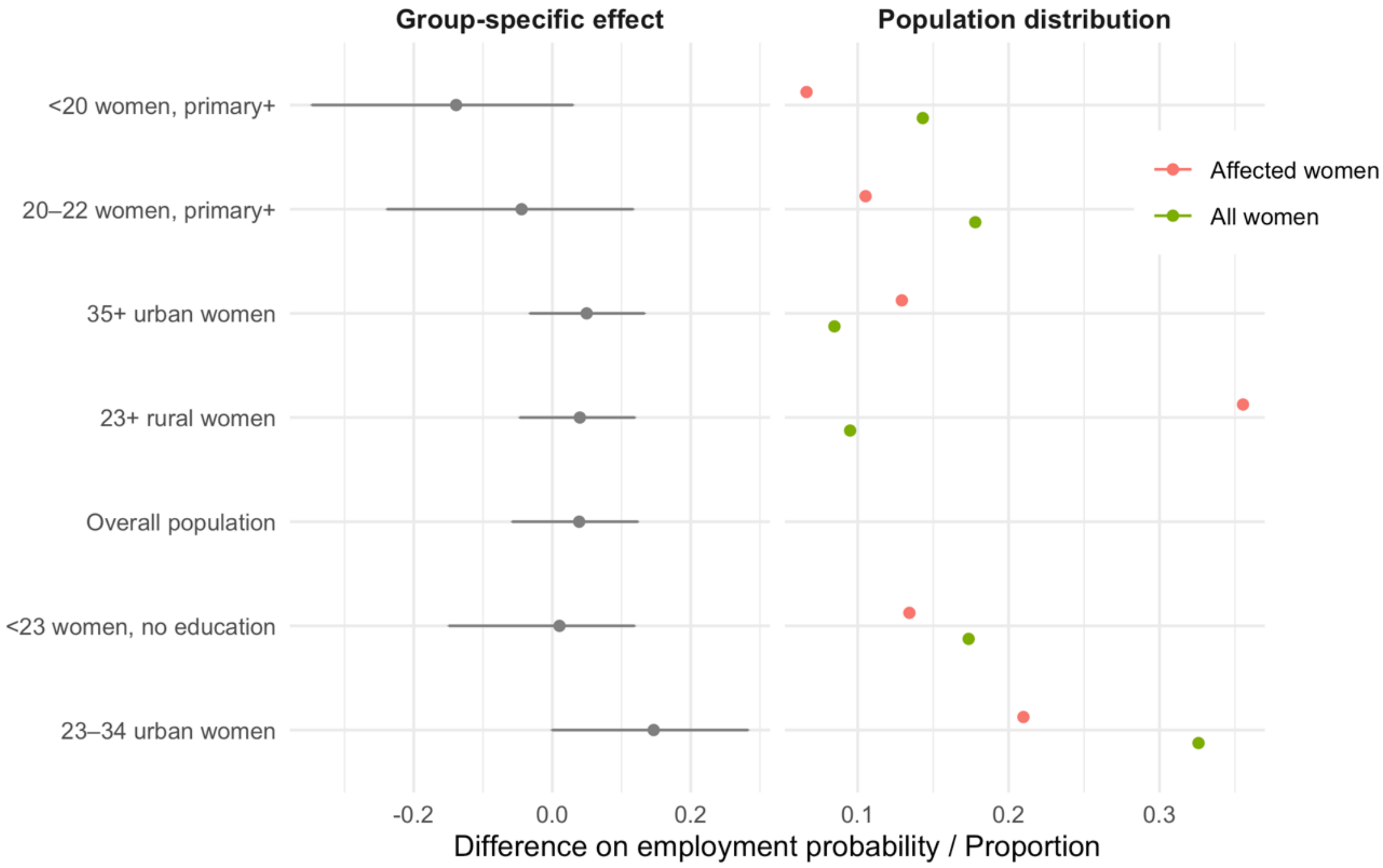

***Figure A 2 Senegal****: Employment effects by subgroup and distribution of the subgroups among women affected by infecundity and all women. Lines represent 90% credible intervals.*

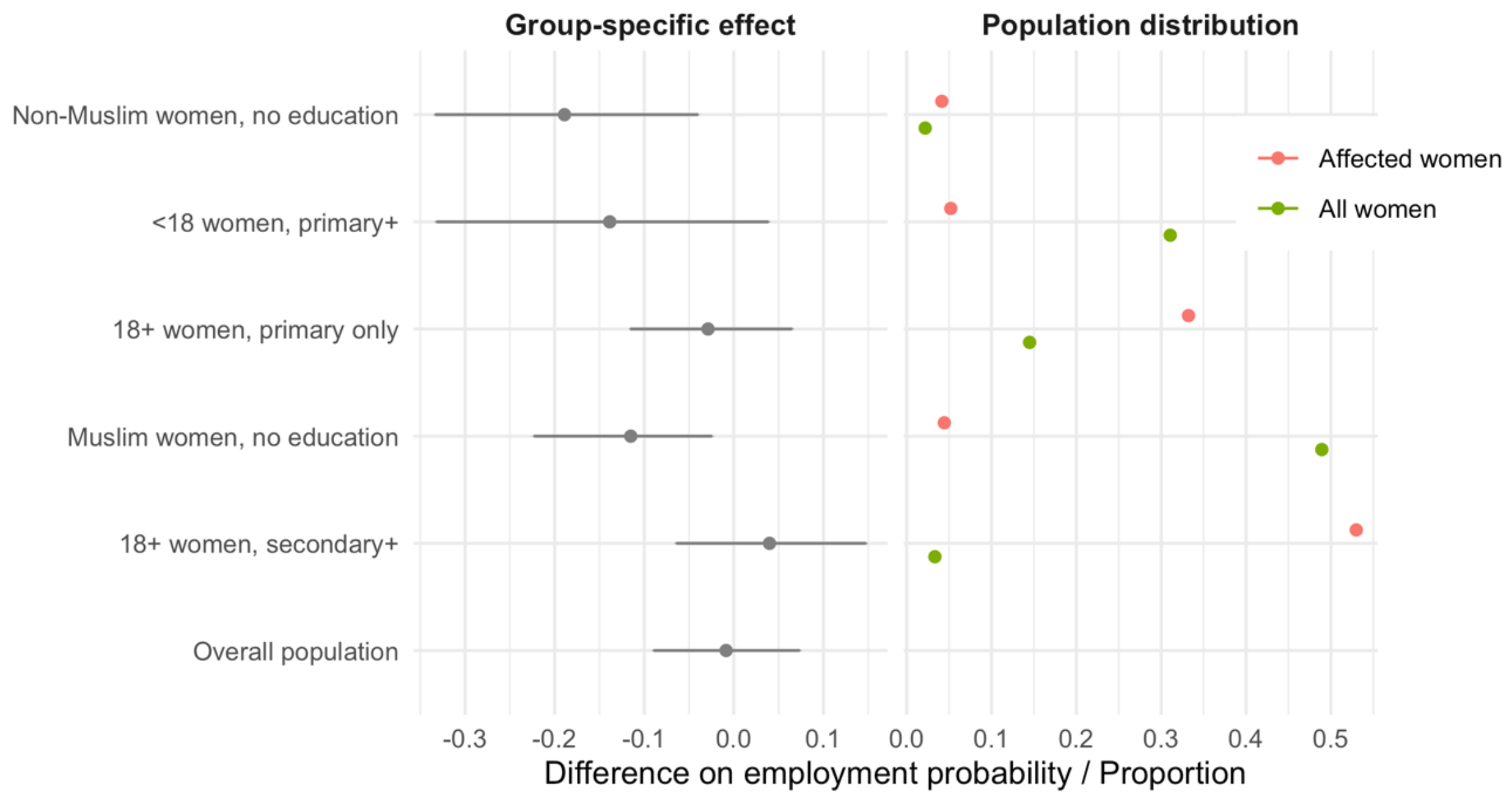

***Figure A 3 Kenya****: Employment effects by subgroup and distribution of the subgroups among women affected by infecundity and all women. Lines represent 90% credible intervals.*